\documentclass[review]{elsarticle}

\usepackage{lineno,hyperref}
\modulolinenumbers[5]

\journal{Elsevier}
\usepackage{calc}
\usepackage{tikz}
\usetikzlibrary{decorations.pathreplacing}
\usepackage{pgfplots}
\usepgfplotslibrary{polar}
\pgfplotsset{compat=newest}
\usetikzlibrary{decorations.markings}
\tikzstyle{vertex}=[circle, draw, inner sep=0pt, minimum size=6pt]
\newcommand{\vertex}{\node[vertex]}
\tikzstyle{vertex}=[circle, draw, inner sep=0pt, minimum size=3pt]
\usepackage{graphicx}
\usepackage{amsmath}
\usepackage{amsfonts}
\usepackage{xcolor}
\usepackage{amssymb}
\usepackage{graphicx}
\usepackage{tikz}
\usetikzlibrary{circuits.logic.US}
\usetikzlibrary{positioning}
\usetikzlibrary{matrix}
\usepackage{graphicx}
\usepackage{amsmath}
\newcommand{\boundellipse}[3]% center, xdim, ydim
{(#1) ellipse (#2 and #3)
}
\definecolor{magenta}{rgb}{0.8, 0.0, 0.8}
\definecolor{cyan}{rgb}{0.0, 1.0, 1.0}
\definecolor{blue1}{rgb}{0.1, 0.6, 0.01}
\definecolor{blue}{rgb}{0.10, 0.50, 1}
\definecolor{brown}{rgb}{0.65, 0.16, 0.16}

\usepackage{tikz}
\usetikzlibrary{decorations.pathreplacing}
\usetikzlibrary{positioning,shapes}
\usetikzlibrary{petri}
\usetikzlibrary{shapes.geometric}
\usetikzlibrary{decorations.markings}
\tikzstyle{vertex}=[circle, draw, inner sep=0pt, minimum size=3pt]
\usetikzlibrary{shapes.geometric}
\tikzstyle{vertex}=[circle, draw, inner sep=0pt, minimum size=3pt]
\tikzstyle{triangle}=[draw, shape=regular polygon, regular polygon sides=3,draw,inner sep=0pt,minimum
size=0.3cm]
\tikzstyle{square}=[draw, shape=regular polygon, regular polygon sides=4,draw,inner sep=0pt,minimum
size=0.225cm]
\usepackage{picture,xcolor}
\usepackage{graphicx}

\usepackage{graphicx}
\usepackage{amsthm}
\usepackage{amsfonts}
\usepackage{xcolor}
\usepackage{amssymb}
\usepackage{marvosym}
\usepackage{graphicx}
\usepackage{tikz}
\usetikzlibrary{circuits.logic.US}
\usetikzlibrary{positioning}
\usetikzlibrary{matrix}
\usepackage{graphicx}
\usepackage{amsmath}
\definecolor{green1}{rgb}{0.1, 0.6, 0.01}
\definecolor{green}{rgb}{0.11, 0.35, 0.02}
\definecolor{brown}{rgb}{0.65, 0.16, 0.16}
\definecolor{magenta}{rgb}{0.8, 0.0, 0.8}
\definecolor{cyan}{rgb}{0.0, 1.0, 1.0}
\definecolor{green1}{rgb}{0.1, 0.6, 0.01}
\definecolor{green}{rgb}{0.11, 0.35, 0.02}
\definecolor{brown}{rgb}{0.65, 0.16, 0.16}
\definecolor{battleshipgrey}{rgb}{0.52, 0.52, 0.51}
\definecolor{babyblue}{rgb}{0.54, 0.81, 0.94}
\definecolor{brightgreen}{rgb}{0.4, 1.0, 0.0}
\definecolor{dimgray}{rgb}{0.41, 0.41, 0.41}
\newtheorem{theorem}{Theorem}

\newtheorem{lemma}[theorem]{Lemma}
\newtheorem{definition}{Definition}

\newtheorem{claim}{Claim}
\newtheorem{example}{Example}
\newtheorem{corollary}{Corollary}
\begin{document}

\begin{frontmatter}
%\title{The Small Set Vertex Expansion Problem\tnoteref{mytitlenote}}

%\title{Elsevier \LaTeX\ template}
%\tnotetext[mytitlenote]{A preliminary version of this paper appeared  in the proceedings of the 14th Annual International Conference on Combinatorial Optimization and Applications  (COCOA) 2020.
%}
\title{Parameterized Complexity of Locally Minimal Defensive Alliances
\tnoteref{mytitlenote}}
\tnotetext[mytitlenote]{A preliminary version has been accepted for publication in the proceedings of the 7th Annual International Conference on Algorithms and Discrete Applied Mathematics (CALDAM) 2021.
}

%\title{Further Parameterized Algorithms for the $\mathcal{F}$-Free Edge Deletion Problem\tnoteref{mytitlenote}}\tnotetext[mytitlenote]{A preliminary version of this paper appeared on the arxiv under the name “Edge deletion to restrict the size of an epidemic”  \cite{DBLP:journals/corr/abs-2102-06068}. }
\author{Ajinkya Gaikwad}
\ead{ajinkya.gaikwad@students.iiserpune.ac.in}

\address{Indian Institute of Science Education and Research, Pune, India}

\author{Soumen Maity\corref{mycorrespondingauthor}}
\cortext[mycorrespondingauthor]{Corresponding author}
\ead{soumen@iiserpune.ac.in}

\address{Indian Institute of Science Education and Research, Pune, India}

\author{Shuvam Kant Tripathi}
\ead{tripathi.shuvamkant@students.iiserpune.ac.in}
\address{Indian Institute of Science Education and Research, Pune, India}

\begin{abstract}
%The {\sc Defensive Alliance} problem has been studied extensively during the last twenty years. 
A set $S$ of vertices of a graph is a defensive alliance if, for each element of $S$, 
the majority of its neighbours is in $S$.
We consider the notion of local minimality in this paper. We are interested in locally 
 minimal defensive alliance of maximum size.
 %We also look at connected version of defensive alliance.
 This problem is known to be NP-hard  but its parameterized complexity remains open 
until now.  We enhance our understanding of the problem from the 
viewpoint of parameterized complexity.
The main results of the paper are the following:
(1)   {\sc Locally Minimal Defensive Alliance}  is NP-complete, even when restricted to planar graphs,
(2) a randomized FPT algorithm for {\sc Exact Connected Locally Minimal Defensive Alliance} parameterized by solution size, (3) {\sc Locally Minimal Defensive Alliance} is 
fixed-parameter tractable (FPT) when parametrized by neighbourhood diversity,
(4) {\sc Locally Minimal Defensive Alliance}  parameterized by treewidth is W[1]-hard and thus not FPT (unless FPT=W[1]),
(5)  {\sc Locally Minimal Defensive Alliance}  can be solved in polynomial time for graphs of bounded treewidth.

\end{abstract}


\end{frontmatter}

\nolinenumbers
\section{Introduction}
%During the last 20 years, the {\sc Defensive Alliance} problem has been studied extensively. 
%A defensive alliance in an undirected graph is a set of vertices with the property that each vertex  has at least as many neighbours  in the alliance (including itself) as neighbours outside the alliance.  
%In 2004, Kristiansen, Hedetniemi, and Hedetniemi \cite{kris} introduced defensive, offensive, and powerful alliances, and these concepts were further studied by Shafique \cite{HassanShafique2004PartitioningAG} and other authors \cite{small,BAZGAN2019111,Cami2006OnTC,SIGARRETA20061345,ROD,SIGARRETA20091687,SIGA,Enciso2009AlliancesIG,BLIEM2018334,Fernau,FERNAU2009177,Lindsay}. 
%In this paper, we will focus on defensive alliances. 
Throughout history, humans have formed communities, guilds, faiths etc in the hope of coming together with a group of people having similar requirements, visions and goals. Their reasons to do so, usually rest on the fact that any group with common interests often provides added mutual benefits to the union in fields of trade, culture, defense, etc as compared to the individual. Such activities are commonly seen in the present day, in areas of geo-politics, cultures, trades, economics, unions etc and are popularly termed as \emph{alliances}. 
Based on the structure, formation and goals of an alliance, many variations of the problem exist in graph theory. 
A defensive alliance is usually formed with the aim of defending its members against non-members, and hence it is natural to ask that each member of the alliance should have more friends within the alliance (including oneself) than outside.
 Similarly, an offensive alliance is formed with the inverse goal of offending or attacking non-members of the alliance. 
 It is known that the problems of finding small 
defensive and offensive alliances are NP-complete. We enhance our understanding of the problems from the 
viewpoint of parameterized complexity.
%  \emph{Strong} versions of the above problems do not consider the self to be a friend, and the \emph{minimal} versions try to find those alliances which lose the required property in the absence of any member.
 \par In 2004, Kristiansen, Hedetniemi, and Hedetniemi \cite{kris}
introduced defensive, offensive, and powerful alliances, and these concepts were 
further studied by Shafique \cite{HassanShafique2004PartitioningAG}
and other authors \cite{small,BAZGAN2019111,Cami2006OnTC,SIGARRETA20061345,ROD,SIGARRETA20091687,SIGA,Enciso2009AlliancesIG,BLIEM2018334,Fernau,FERNAU2009177,Lindsay}. 
%In this paper, we will focus on defensive alliances. 
%A defensive alliance is {\it strong} if each vertex has at least as many neighbours in the alliance (not counting itself) as outside the alliance. 
The theory of alliances in graphs has been studied 
intensively \cite{frick,Cami2006OnTC,10.5614/ejgta.2014.2.1.7} both from a combinatorial and from a computational perspective. 
As mentioned in \cite{BAZGAN2019111}, the 
focus has been mostly on finding small alliances, although studying large
alliances does not only make a lot of sense from the original motivation of these notions, 
but was actually also delineated in the very first papers on alliances \cite{kris}.
\par Note that being a defensive alliance is not a hereditary property, that is, a superset or subset of a
defensive alliance is not necessarily a  defensive alliance.  Shafique \cite{HassanShafique2004PartitioningAG} called 
an alliance a {\it locally minimal alliance} if the set obtained by removing any vertex of the
alliance is not an alliance. Bazgan  et al. \cite{BAZGAN2019111} considered another notion 
of alliance that they called a {\it globally minimal alliance} which has the property that 
no proper subset is an alliance. In this paper we are interested in finding locally minimal 
alliances of size at least $k$.  Bazgan et al. \cite{BAZGAN2019111} proved that deciding if a graph contains a  locally minimal defensive alliance of size at least $k$ is NP-complete, even when restricted to 
 bipartite graphs with average degree less than 5.6.  Clearly, the motivation is that big communities where every member
still matters somehow are of more interest than really small communities. Also, there is a 
general mathematical interest in such type of problems, see \cite{Fernau202248,Manlove1998MinimaximalAM}.
\par Motivation for maximum-minimal / minimum-maximal problems can be given as follows.  Many local search heuristics for NP-hard optimisation problems can be modelled by defining a partial order on the feasible solutions that are iteratively improved upon.  Once we reach a solution that is minimal or maximal with respect to the given partial order, the local search process terminates.
For example, for {\sc Independent Set}, find a maximal independent set by selecting an unmarked vertex $v$, marking $v$ and all of its neighbours, iterating this approach until no unmarked vertex remains.  The partial order is the strict subset relation.
 For {\sc Chromatic Number}, starting from an arbitrary graph colouring, try to ``improve” it by using the following recolouring strategy.  Try to find some colour $c$ such that each vertex of colour $c$ can be recoloured by one of the remaining colours, and iterate this process.  The partial order here is one of partition refinement.  See Chapter 2 of  PhD thesis \cite{Manlove1998MinimaximalAM} for further details.
 We assume a relationship between the partial order used and the measure function of the source optimisation problem called “partial order measure monotonicity” – that is, every time we obtain a “local” improvement relative to the partial order, we obtain a “global” improvement (e.g., size of independent set goes up by 1, number of colours used goes down by 1) relative to the measure function.
 
The minimum-maximal or maximum-minimal problem then corresponds to the worst-case behaviour of such a local search heuristic.  These problems can be interesting in their own right.  Note that  {\sc Minimum Maximal Independent Set} is the same as the extensively studied  {\sc Minimum Independent Dominating Set} \cite{GODDARD2013839}.  Similarly
{\sc Maximum Minimal Chromatic Number} is the same as a completely new problem called 
{\sc B-Chromatic Number}, which ended up getting a lot of attention in the literature
\cite{JAKOVAC2018184,Manlove1998MinimaximalAM}.
 
Also note that enumerating all minimal (resp. maximal) solutions can be a useful strategy in relation to solving the original minimization (resp. maximization) problem.
The enumeration problem asks to enumerate all  minimal (resp. maximal) solutions 
for a given input instance. The existence of an enumeration algorithm which runs in time $O^*(c^n)$  implies the existence of an 
$O^*(c^n)$ for the source optimization problem. This follows from the fact that the solution to the minimization (resp. maximization)  problem can be obtained by enumerating all minimal (resp. maximal) solutions and then look for the smallest (resp. largest)  of the enumerated solutions. See \cite{khosravianghadikolaei:tel-03220653,Fernau202248} for details. 

%the PhD thesis of  Mehdi Khosravian  worked on this aspect as part of his PhD thesis and I recommend checking out his work at https://dblp.org/pid/210/3512.html and https://theses.hal.science/tel-03220653.

\section{Basic Notations} 
Throughout this article, $G=(V,E)$ denotes a finite, simple and undirected graph of order $|V|=n$. 
The {\it (open) neighbourhood} $N_G(v)$ of a vertex 
$v\in V(G)$ is the set $\{u~|~(u,v)\in E(G)\}$. The {\it closed neighbourhood} $N_G[v]$ of a vertex $v\in V(G)$ is the set
$\{v\} \cup N_G(v)$.  The {\it degree} of $v\in V(G)$ is $|N_G(v)|$ and denoted by $d_G(v)$.
The subgraph induced by  $S\subseteq V(G)$ is denoted by $G[S]$. For a non-empty subset $S\subseteq V$ and a vertex $v\in V$, 
$d_S(v)$ denotes the number of neighbours that $v$ has in the vertex set $S$.
%The subgraph induced by  $S\subseteq V(G)$ is denoted by $G[S]$. We use $d_S(v)$ to denote the number of neighbours the degree of vertex $v$ in $G[S]$. 
The complement of the vertex set $S$ in $V$ is denoted by $S^c$.
\begin{definition}\rm
A non-empty set $S\subseteq V$ is a \emph{defensive alliance} in $G$ if for each $v\in S$, 
   $d_S(v)+1\geq d_{S^c}(v)$. 
\end{definition}\noindent 
\noindent We often use the terms \emph{defenders}  and \emph{attackers} 
of an element $v$ of a defensive alliance  $S$. By these we mean the sets  
$N[v] \cap S$ and 
$N[v] \setminus S$ respectively. Thus, including itself, $v$ has $|N[v] \cap S|=d_S(v)+1$ defenders and 
$v$ has $|N[v] \setminus S|= d_{S^c}(v)$ attackers in $G$. A vertex $v\in S$ is said to be  \emph{protected} if $d_S(v)+1\geq d_{S^c}(v)$. A vertex $v\in S$ is said to be  \emph{unprotected} if $d_S(v)+1< d_{S^c}(v)$. A set $S\subseteq V$ is a  defensive alliance if every vertex
in $S$ is  protected.

\begin{definition}\rm
A vertex $v\in S$ is said to be {\it marginally protected} 
if it becomes unprotected when any of its neighbours in $S$ is moved from $S$ to $V\setminus S$.
A vertex $v\in S$ is said to be {\it overprotected} if it remains protected
even when any of its neighbours is moved from $S$ to $V\setminus S$. 
\end{definition}

%A  defensive alliance is connected if the subgraph induced by $S$ is connected. 
\begin{definition}\rm \cite{BAZGAN2019111}
A defensive alliance $S$ is called a {\it locally minimal defensive alliance}
if for any $v\in S$, $S\setminus \{v\}$ is not a defensive alliance.
\end{definition}

\noindent \emph{It is important to note  that $S$ is a locally minimal  defensive alliance in $G$ 
 if and only if for every vertex $v \in S$, at least one of its neighbours in $S$ is marginally protected.}
 \begin{definition}\rm \cite{BAZGAN2019111}
A defensive alliance $S$ is called a {\it globally minimal defensive alliance}  if no proper 
subset is a defensive alliance.
\end{definition}
\begin{example}\rm
Consider the tree  in Figure \ref{treeLMDA}. It has a locally minimal defensive alliance 
 $S_1=\{2,3,4,5,6,7,9,11,13,15\}$ 
of size 10 and  and a globally minimal defensive alliance $S_2=\{1,2,3\}$  of size 3.
Note that, including itself, vertex $2$ has two defenders and it has two attackers; so vertex 2 is  marginally protected. Similarly, including itself, vertex 7 has two defenders and one attacker; so vertex $7$ is marginally protected.   It is easy to see that  for every vertex $v \in S_1$, at least one of its neighbours in $S_1$ is marginally protected. Hence, $S_1$ is a locally minimal defensive alliance. Note that 
$S_1$ is not a globally minimal defensive alliance as $\{2,7\}$, a proper subset of $S_1$, is also a defensive alliance. On the other hand, no proper subset of $S_2$ is a 
defensive alliance, hence $S_2$ is a globally minimal defensive alliance. It may be noted that every  globally minimal defensive alliance is also a locally minimal defensive alliance but the converse is not true. 
\end{example}

\begin{figure}[ht]
     \centering
  \[  \begin{tikzpicture}[scale=0.5]
\node[vertex](v1)at(6,0){$1$};
\node[vertex,fill=yellow](v2)at(0,-2){2};
\node[vertex,fill=yellow](v3)at(3,-2){3};
\node[vertex,fill=yellow](v4)at(6,-2){4};
\node[vertex,fill=yellow](v5)at(9,-2){5};
\node[vertex,fill=yellow](v6)at(12,-2){6};
\node[vertex,fill=yellow](v7)at(-1,-4){7};
\node[vertex](v8)at(1,-4){8};
\node[vertex,fill=yellow](v9)at(2,-4){9};
\node[vertex](v10)at(4,-4){10};
\node[vertex,fill=yellow](v11)at(5,-4){11};
\node[vertex](v12)at(7,-4){12};
\node[vertex,fill=yellow](v13)at(8,-4){13};
\node[vertex](v14)at(10,-4){14};
\node[vertex,fill=yellow](v15)at(11,-4){15};
\node[vertex](v16)at(13,-4){16};
\node[vertex](v17)at(-1,-6){17};
\node[vertex](v18)at(1,-6){18};
\node[vertex](v19)at(2,-6){19};
\node[vertex](v20)at(4,-6){20};
\node[vertex](v21)at(5,-6){21};
\node[vertex](v22)at(7,-6){22};
\node[vertex](v23)at(8,-6){23};
\node[vertex](v24)at(10,-6){24};
\node[vertex](v25)at(11,-6){25};
\node[vertex](v26)at(13,-6){26};
%layer one edges
\path
    (v1) edge (v2)
    (v1) edge (v3)
    (v1) edge (v4)
    (v1) edge (v5)
    (v1) edge (v6)
    (v2) edge (v7)
    (v2) edge (v8)
    (v3) edge (v9)
    (v3) edge (v10)
    (v4) edge (v11)
    (v4) edge (v12)
    (v5) edge (v13)
    (v5) edge (v14)
    (v6) edge (v15)
    (v6) edge (v16)
    (v7) edge (v17)
    (v8) edge (v18)
    (v9) edge (v19)
    (v10) edge (v20)
    (v11) edge (v21)
    (v12) edge (v22)
    (v13) edge (v23)
    (v14) edge (v24)
    (v15) edge (v25)
    (v16) edge (v26);
\end{tikzpicture}\]
\caption{ A tree. }
  \label{treeLMDA}
 \end{figure}
A graph is said to be {\it connected} if there is a path between every pair of vertex.
A locally minimal defensive alliance $S$ is called a  {\it connected locally minimal defensive alliance} if the subgraph induced by 
$S$ is connected. 
%A connected defensive alliance $S$ is called a {\it  connected locally minimal defensive alliance} if for any $v\in S$, $S\setminus \{v\}$ is not a connected defensive alliance.  
Notice that
any globally  minimal defensive alliance is always connected. 
%As introduced in \cite{BAZGAN2019111}, we use 
%$A_L(G)$ for the cardinality of the largest locally minimal defensive alliance in a graph $G$, and 
%$A(G)$ for the cardinality of the largest globally minimal defensive alliance in a graph $G$. 
In this paper, we study  {\sc Locally Minimal Defensive Alliance} and {\sc Exact Connected Locally Minimal Defensive Alliance}. 
 We define the problems as follows:
    \vspace{3mm}
    \\
    \fbox
    {\begin{minipage}{33.7em}\label{FFVS0 }
       {\sc Locally Minimal Defensive Alliance}\\
        \noindent{\bf Input:} An undirected graph $G=(V,E)$ and an  integer $k \leq |V(G)|$.
    
        \noindent{\bf Question:} Is there a locally minimal defensive alliance $S\subseteq V(G)$ such that 
        $|S|\geq  k $?
    \end{minipage} }
  \vspace{3mm}   
   \vspace{3mm}
    \\
    \fbox
    {\begin{minipage}{33.7em}\label{FFVS1 }
       {\sc Exact Connected  Locally Minimal Defensive Alliance}\\
        \noindent{\bf Input:} An undirected graph $G=(V,E)$ and an  integer $k \leq |V(G)|$.
    
        \noindent{\bf Question:} Is there a  connected locally minimal  defensive alliance $S\subseteq V(G)$ such that 
        $|S|= k $?
    \end{minipage} }
  \vspace{3mm}  
  
  %\noindent While {\sc Connected Locally Minimal Defensive Alliance}  asks for connected locally minimal defensive alliances of size at most $k$, we also consider  {\sc Exact Connected Locally Minimal Defensive Alliance}  that concerns defensive alliances of size exactly $k$.  Given a graph $G = (V, E)$, we also study {\sc Connected Locally Minimal Strong Defensive Alliance}, where the goal is to find a largest connected locally minimal strong defensive alliance  in $G$. \\

  The graph
parameter that we explicitly use in this paper is  treewidth.
We  review the concept of a tree decomposition, introduced by Robertson and Seymour in \cite{Neil}.
Treewidth is a  measure of how “tree-like” the graph is.
\begin{definition}\rm \cite{Downey} A {\it tree decomposition} of a graph $G=(V,E)$  is a tree $T$ together with a 
collection of subsets $X_t$ (called bags) of $V$ labeled by the vertices $t$ of $T$ such that 
$\bigcup_{t\in T}X_t=V $ and (1) and (2) below hold:
\begin{enumerate}
			\item For every edge $uv \in E(G)$, there  is some $t$ such that $\{u,v\}\subseteq X_t$.
			\item  (Interpolation Property) If $t$ is a vertex on the unique path in $T$ from $t_1$ to $t_2$, then 
			$X_{t_1}\cap X_{t_2}\subseteq X_t$.
		\end{enumerate}
	\end{definition}
	
%\noindent It is important to note that a graph may have several different tree decomposition.
%	Similarly, the same tree decomposition can be valid for several different graphs. 
%	Every graph has a trivial tree decomposition for which $T$ has only one vertex including all
%	of $V$. However, this is not effective for the purpose of solving problems. 
\begin{definition}\rm \cite{Downey} The {\it width} of a tree decomposition is
the maximum value of $|X_t|-1 $ taken over all the vertices $t$ of the tree $T$ of the decomposition.
The treewidth $tw(G)$ of a graph $G$  is the  minimum width among all possible tree decomposition of $G$.
\end{definition}

%The reason for subtracting 1 in the above definition for width is so that we can define forests as having treewidth 1.
\noindent A special type of tree decomposition, known as a {\it nice tree decomposition} was 
introduced by Kloks \cite{Kloks1994TreewidthCA}.  The nodes in such a decomposition can be partitioned into four types:

\begin{definition}\rm \cite{Kloks1994TreewidthCA} A tree decomposition is said to be {\it nice tree decomposition} if the following conditions are satisfied:
	\begin{enumerate}
		\item All bags correspond to leaves are empty. One of the leaves is considered as root node $r$. Thus $X_r=\emptyset$ and $X_{l}=\emptyset$ for each leaf $l$. 
		\item There are three types of non-leaf nodes:
		\begin{itemize}
		 \item {\bf Introduce node:} a node $t$ with exactly one child $t^{\prime}$
		such that $X_{t}=X_{t^{\prime}} \cup \{v\}$ for some $v \notin X_{t^{\prime}}$;
		we say that $v$ is {\it introduced} at $t$.
		\item {\bf Forget node:} a node $t$ with exactly one child $t^{\prime}$
		such that $X_{t}=X_{t^{\prime}} \setminus \{w\}$ for some $w \in X_{t^{\prime}}$;
		we say that $w$ is {\it forgotten}  at $t$.
		\item {\bf Join node:} a node with two children $t_1$ and $t_2$ such that $X_t=X_{t_1}=X_{t_2}$.
		\end{itemize}
	\end{enumerate}
\end{definition}

\noindent Note that, by the third property of tree decomposition, a vertex $v\in V(G)$ may be introduced several time, but each vertex is forgotten  only once. 
%To control introduction of edges, sometimes one more type of node is considered in nice tree decomposition called introduce edge node. An {\it introduce edge node} is a node $t$, labeled with edge $uv \in E(G)$, such that $u,v\in X_{t}$ and $X_{t}=X_{t^{\prime}}$, where $t^{\prime}$ is the only child of $t$. We say that node $t$ introduces edge $uv$. We additionally require that every edge of $E(G)$ is introduced exactly once in the whole decomposition.
It is known that if a graph	$G$ admits a tree decomposition of width at most {\tt tw}, then it also admits	a nice tree decomposition of width at most {\tt tw}, that has at most $O(n\cdot {\tt tw})$ nodes \cite{marekcygan}.\\

\par We recall the definitions of treedepth, vertex cover and feedback vertex set which are 
used in Section \ref{W[1]-section}. A rooted forest is a disjoint union of rooted trees. Given a rooted forest $F$, its \emph{transitive closure} is a graph $H$ in which $V(H)$ contains all the nodes of the rooted forest, and $E(H)$ contain an edge between two vertices only if those two vertices form an ancestor-descendant pair in the forest $F$.

   \begin{definition}
        {\rm  The {\it treedepth} of a graph $G$ is the minimum height of a rooted forest $F$ whose transitive closure contains the graph $G$. It is denoted by $td(G)$.}
    \end{definition}

\begin{definition}\rm 
A set $S \subseteq V(G)$ is a \emph{vertex cover} of $G=(V,E)$ if  each edge in $E$ has at least one  endpoint in $S$.   The \emph{size} of a smallest vertex cover of  $G$ is the \emph{vertex cover number} of $G$.
\end{definition}

\begin{definition}\rm 
 A \emph{feedback vertex set}  of a graph  $G$ is a set of vertices whose removal leaves $G$ without cycles. The minimum size of a feedback vertex set in $G$ is the {\it feedback vertex set number} of $G$, denoted by ${\sf fvs}(G)$.
\end{definition}

\noindent We now recall the definition of Iverson bracket that will be used in the proof of Theorem  \ref{theoremnd1}.  Theorem  \ref{theoremnd1} proves that {\sc Locally Minimal 
Defensive Alliance} is fixed-parameter tractable 
when parameterized by the neighbourhood diversity of the input graph.
\begin{definition} \label{Iverson}\rm 
      Let $S$ be a mathematical statement, then the \emph{Iverson bracket}
      is defined by
      $$
[S]=
\begin{cases}
1 &  \mbox{if $S$ is true}\\
0 & \mbox{if $S$ is false}\\
\end{cases}
$$
and corresponds to the so-called characteristic function or indicator function.    
    \end{definition}  \subsection{Parameterized Complexity}  
A \emph{parameterized problem} is a language $L\subseteq \Sigma^{\star} \times \mathbb{N}$,
where $\Sigma $
is a fixed, finite alphabet. For an instance $(x,k)\in \Sigma^{\star} \times \mathbb{N}$,
$k$ is called the \emph{parameter}. A parameterized problem $\mathcal{P}$ is \emph{fixed-parameter tractable} (FPT in short) if a given instance $(x,k)$ can be solved in time $f(k)\cdot {|(x,k)|}^c$
 where $f$ is some (usually computable) function, and $c$ is a constant. Parameterized complexity classes are defined with respect to {\it fpt-reducibility}. A parameterized problem $\mathcal{P}$ is {\it fpt-reducible} to 
 $\mathcal{Q}$ if in time  $f(k) \cdot {|(x,k)|}^c$, one can transform an instance 
 $(x, k)$ of $\mathcal{P}$ into an instance $(x', k')$ of $Q$ such that $(x, k) \in \mathcal{P}$
 if and only if  $(x',k') \in \mathcal{Q}$, and $k'\leq g(k)$, where $f$ and $g$ are computable functions depending only on $k$. Owing to the definition, if $\mathcal{P}$ {\it fpt-reduces} to $\mathcal{Q}$ and $\mathcal{Q}$ is fixed-parameter tractable then $\mathcal{P}$ is fixed-parameter tractable as well.

 What makes the theory more interesting is a hierarchy of intractable parameterized problem classes above FPT which helps in distinguishing those problems that are not fixed parameter tractable. 
 Central to parameterized complexity is the following hierarchy of complexity classes, defined by the closure of canonical problems under {\it fpt-reductions}: FPT $\subseteq$ W[1] $\subseteq$ W[2] $\subseteq \ldots \subseteq $ XP. All inclusions are believed to be strict. In particular, FPT $\neq$  W[1] under the Exponential Time Hypothesis \cite{Paturi}.
The class W[1] is the analog of NP in parameterized complexity. A major goal in parameterized complexity is to distinguish between parameterized problems which are in FPT
 and those which are \emph{W[1]-hard}, i.e., those to which every problem in W[1] is \emph{fpt-reducible}. There are many problems shown to be complete for W[1], or equivalently \emph{W[1]-complete}, including the {\sc MultiColored Clique} (MCC) problem \cite{Downey}.
 We refer to \cite{marekcygan,Downey} for further details on parameterized complexity. \\

\subsection{Our results}    
In this paper, we study  {\sc Locally Minimal Defensive Alliance} and {\sc Exact Connected Locally Minimal Defensive Alliance} mainly from the parameterized complexity point of view. We show  both tractability and intractability results. Our results are the following:
\begin{itemize}
% \item   {\sc Connected Locally Minimal Strong Defensive Alliance}  is polynomial time solvable on trees.
\item {\sc Locally Minimal Defensive Alliance}  is NP-complete, even when restricted to planar graphs.
\item {\sc Locally Minimal Defensive Alliance}  is fixed-parameter tractable (FPT) when parameterized 
by neighbourhood diversity.
\item {\sc Locally Minimal Defensive Alliance} parameterized by treewidth is W[1]-hard and thus not FPT (unless
FPT=W[1]).
 \item   {\sc Locally Minimal Defensive Alliance} problem is polynomial time solvable 
for graphs with bounded treewidth. That is, the problem can be solved in XP-time when parameterized by treewidth.
\item Finally, we give a randomized FPT algorithm for {\sc Exact Connected 
Locally Minimal Defensive Alliance} when parameterized by the solution size $k$.
\end{itemize} 

\begin{figure}[ht]
     \centering
 \begin{tikzpicture}[%
    auto,
    block/.style={
      rectangle,
      draw=black,
      thick,
      text width=3em,
      align=center,
      rounded corners,
      minimum height=1.5em
    }
]
    \draw (0,0) node[block] (vc) {\color{blue} vc$^*$}
          (2,1.5) node[block] (nd) {\color{blue} nd$^*$}
          (4,1.5) node[block] (tc) {\color{black} tc}
          (-3,1.5) node[block] (vi) {\color{black} vi}
          (-3,2.5) node[block] (td) { \color{red}  td$^*$}
          (-2,3.5) node[block] (fvs) {\color{red} fvs$^*$}
          (-4,3.5) node[block] (pw) { \color{red} pw$^*$}
          (2,3) node[block] (mw) {mw}
          (4,3) node[block] (cvd) { \color{black} cvd}
          (-3,4.5) node[block] (tw) {\color{red} tw$^*$}
          (-1,5.5) node[block] (cw) {\color{red} cw$^*$};
          
\draw[->, thick] (vc)--(vi); 
\draw[->, thick] (vc)--(nd); 
\draw[->, thick] (vc)--(tc); 
\draw[->, thick] (vi)--(td); 
\draw[->, thick] (td)--(pw); 
\draw[->, thick] (vc)--(fvs); 
\draw[->, thick] (nd)--(mw); 
\draw[->, thick] (tc)--(mw); 
\draw[->, thick] (tc)--(cvd); 
\draw[->, thick] (pw)--(tw); 
\draw[->, thick] (fvs)--(tw); 
\draw[->, thick] (tw)--(cw); 
\draw[->, thick] (mw)--(cw); 
\draw[->, thick] (cvd)--(cw);

%\draw[green, thick, dashed] (-4,-.5)--(5,-0.5); 
%\draw[green, thick, dashed] (-4,1.8)--(5,1.8); 
%\draw[green, thick, dashed] (-4,1.8)--(-4,-.5); 
%\draw[green, thick, dashed] (5,-.5)--(5,1.8); 

%\draw[red, thick, dashed] (-5,2)--(0,2); 
%\draw[red, thick, dashed] (-5,6)--(0,6); 
%\draw[red, thick, dashed] (-5,2)--(-5,6); 
%\draw[red, thick, dashed] (0,2)--(0,6); 

%\draw[blue, thick, dashed] (1,2)--(5,2); 
%\draw[blue, thick, dashed] (1,6)--(5,6); 
%\draw[blue, thick, dashed] (1,2)--(1,6); 
%\draw[blue, thick, dashed] (5,2)--(5,6); 

  \end{tikzpicture}

\caption{ Relationship between vertex cover (vc), neighbourhood diversity (nd), twin cover (tc), modular width (mw), cluster vertex deletion number (cvd), feedback vertex set (fvs), pathwidth (pw), treewidth (tw), vertex integrity (vi) and clique width (cw). 
%Arrow indicate generalizations, for example, treewidth generalizes both feedback vertex set and pathwidth.
Note that $A\rightarrow B$ means that there exists a function $f$ such that for 
all graphs, $f(A(G))\geq B(G)$. It also gives an overview of the parameterized complexity landscape for  {\sc Locally Minimal Defensive Alliance}.
The problem is FPT parameterized by blue colored parameters and W[1]-hard 
when parameterized by red colored parameters. The stars highlight parameters that are covered in this paper. The problem remains unsettled when parameterized by mw, vi, cvd and tc. }

     \label{fig:parameters}
 \end{figure}
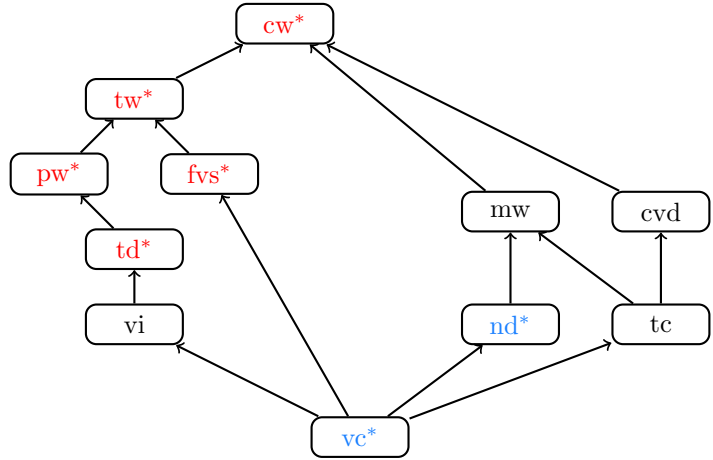

\noindent{\it Known Results:} The decision version for several types of alliances have been shown to be NP-complete. 
 Carvajal et al. \cite{small} proved that deciding if a graph 
 contains a strong defensive alliance of size at most $k$ is NP-hard. 
  The problem of deciding if a graph contains defensive alliance 
  of size at most $k$ is
 NP-complete even when restricted to split, chordal and bipartite graphs \cite{Lindsay}. 
 Bazgan et al. \cite{BAZGAN2019111} proved that deciding if a graph contains a  locally minimal 
 strong defensive alliance of size at least $k$ is NP-complete, even when restricted to 
 bipartite graphs with average degree less than 3.6. Bazgan et al. \cite{BAZGAN2019111} also proved that deciding if a graph contains a connected locally minimal 
 strong defensive alliance or a connected locally minimal 
  defensive alliance of size at least $k$ is NP-complete, even when restricted to 
 bipartite graphs with average degree less than $2+\epsilon$, for any $\epsilon >0$.

\section{{\sc Locally Minimal Defensive Alliance} in planar graphs is NP-complete}
 Bazgan, Fernau and Tuza showed in \cite{BAZGAN2019111} that 
 the problem of  deciding if a graph contains a 
 locally minimal defensive alliance of size at least $k$ for bipartite graphs with average degree less than 5.6 and the problem of 
 deciding if a graph contains a connected locally minimal  defensive alliance of size at least $k$, even for 
 bipartite graphs with average degree less than $2 + \epsilon$, for any $ \epsilon > 0$,
 are NP-complete.
Here we prove that  {\sc Locally Minimal Defensive Alliance} is NP-complete in planar graphs, via 
a reduction from {\sc Minimum Maximal Matching}  in cubic planar graph.  
Yannakakis and Gavril showed in \cite{MMMYann} that the problem of finding a maximal matching of minimum size, 
 is NP-hard in  planar graphs of maximum degree 3.
In \cite{MMMHorton}, Horton and Kilakos obtained the NP-hardness of {\sc Minimum Maximal Matching}
in  planar cubic graphs.

 \begin{theorem}\label{Ffvsthm}
 {\sc Locally Minimal Defensive Alliance}  is NP-complete, even when restricted to planar graphs.
 \end{theorem}
 
\begin{figure}[ht]
 \usetikzlibrary{petri}
\usetikzlibrary{shapes.geometric}
\tikzstyle{square}=[draw]
     \centering
    \[\begin{tikzpicture}[scale=1]
    \tiny[square/.style={draw,regular polygon,regular polygon sides=4,minimum size=1},outer sep=0,inner sep=0]
	%\node[square, draw=black] (a1) at (1.5,6.3) [label=above:] {};
    \tikzstyle{triangle}=[draw, shape=regular polygon, regular polygon sides=3,draw,inner sep=0pt,minimum
size=0.25cm],
	%% Notice in the first vertex is named (v) for the sake of a later edge,
	%% and it also has a label to its left that is the math-mode $v$. 
    %\node[triangle, draw=black, fill=red] (a) at (.5,4.5) [label=above right:${\color{red} a}$] {};
	%% Notice in the first vertex is named (v) for the sake of a later edge,
	%% and it also has a label to its left that is the math-mode $v$. 
	%\vertex (a) at (0.5,4.5) [label=above right:${\color{red} a}$] {};
	%\node[square, draw=black, fill=blue] (a1) at (0.5,5.5) [label=above:${\color{blue} a^{\square}}$] {};
	%\node[square, draw=black] (ha1) at (-0.5,5.2) [label=above:] {};
	\vertex (d) at (-1,0) [label=below:$d$] {}; 
	\vertex (a) at (-1,1) [label=above:$a$] {};
	\vertex (b) at (0,1) [label=above:$b$] {};
	\vertex (c) at (0,0) [label=below:$c$] {};
	
	\vertex (dn) at (4,-1) [label=above:$d$] {}; 
	\vertex (an) at (4,2) [label=above:$a$] {};
	\node[] (z1) at (4,2.5) [label=above: $A$ ]{};
	\vertex (bn) at (4,1) [label=above:$b$] {};
	\vertex (cn) at (4,0) [label=above:$c$] {};
	\vertex (dn') at (3,-1) [label=above:$d^{\square}$] {}; 
	\vertex (an') at (3,2) [label=above:$a^{\square}$] {};
%	\filldraw[color=red!60, fill=red!5, very thick] (2,2) circle (.2);
	\node[] (z1) at (3,2.5) [label=above: $V^{\square}$ ]{};

	\vertex (bn') at (3,1) [label=above:$b^{\square}$] {};
	\vertex (cn') at (3,0) [label=above:$c^{\square}$] {};
	\vertex (ab) at (5.5,3.6) [label=above:$ab$] {};
	
	\node[] (z1) at (5.5,4.3) [label=above: $B$ ]{};
	\vertex (ab_s) at (6.5,3.9) [label=above:$ab^{\square}$] {}; 
	\node[] (z1) at (6.5,4.3) [label=above: $E^{\square}$ ]{};
	
	\vertex (ab'_s) at (6.5,3.3) [label=above:$ab^{'\square}$] {}; 
	
	\vertex (ac) at (5.5,2.4) [label=above:$ac$] {};
	\vertex (ac_s) at (6.5,2.7) [label=above:$ac^{\square}$] {}; 
	\vertex (ac'_s) at (6.5,2.1) [label=above:$ac^{'\square}$] {}; 
	
	\vertex (ad) at (5.5,1.2) [label=above:$ad$] {};
	\vertex (ad_s) at (6.5,1.5) [label=above:$ad^{\square}$] {}; 
	\vertex (ad'_s) at (6.5,0.9) [label=above:$ad^{'\square}$] {};

	\vertex (bc) at (5.5,-0.2) [label=above:$bc$] {};
	\vertex (bc_s) at (6.5,0.1) [label=above:$bc^{\square}$] {}; 
	\vertex (bc'_s) at (6.5,-0.5) [label=above:$bc^{'\square}$] {};

	\vertex (bd) at (5.5,-1.4) [label=above:$bd$] {}; 
	\vertex (bd_s) at (6.5,-1.1) [label=above:$bd^{\square}$] {}; 
	\vertex (bd'_s) at (6.5,-1.7) [label=above:$bd^{'\square}$] {};
	
	\vertex (cd) at (5.5,-2.6) [label=above:$cd$] {};
	\vertex (cd_s) at (6.5,-2.3) [label=above:$cd^{\square}$] {}; 
	\vertex (cd'_s) at (6.5,-2.9) [label=above:$cd^{'\square}$] {};
	
\node[] (z1) at (-0.4,-2.9) [label=below: (a) ]{};	
\node[] (z1) at (5,-2.9) [label=below: (b) ]{};

\node[circle, draw, red, thick,  minimum size=0.4cm] (a0) at (2,2) [label=left:$C_{a^{\square}}$]{};
\node[circle, draw, red, thick,  minimum size=0.4cm] (b0) at (2,1) [label=left:$C_{b^{\square}}$]{};
\node[circle, draw, red, thick,  minimum size=0.4cm] (c0) at (2,0) [label=left:$C_{c^{\square}}$]{};
\node[circle, draw, red, thick,  minimum size=0.4cm] (d0) at (2,-1) [label=left:$C_{d^{\square}}$]{};

\node[circle, draw, thick, red, minimum size=0.4cm] (ab0) at (7.5,3.9) [label=right:$C_{ab^{\square}}$]{};
\node[circle, draw, thick, red, minimum size=0.4cm] (ab00) at (7.5,3.3) [label=right:$C_{ab'^{\square}}$]{};
\node[circle, draw, thick, red, minimum size=0.4cm] (ac0) at (7.5,2.7) [label=right:$C_{ac^{\square}}$]{};
\node[circle, draw, thick, red, minimum size=0.4cm] (ac00) at (7.5,2.1) [label=right:$C_{ac'^{\square}}$]{};
\node[circle, draw, thick,  red, minimum size=0.4cm] (ad0) at (7.5,1.5) [label=right:$C_{ad^{\square}}$]{};
\node[circle, draw, thick, red, minimum size=0.4cm] (ad00) at (7.5,0.9) [label=right:$C_{ad'^{\square}}$]{};
\node[circle, draw, thick, red, minimum size=0.4cm] (bc0) at (7.5,0.1) [label=right:$C_{bc^{\square}}$]{};
\node[circle, draw, thick, red, minimum size=0.4cm] (bc00) at (7.5,-0.5) [label=right:$C_{bc'^{\square}}$]{};
\node[circle, draw, red, thick, minimum size=0.4cm] (bd0) at (7.5,-1.1) [label=right:$C_{bd^{\square}}$]{};
\node[circle, draw, red, thick, minimum size=0.4cm] (bd00) at (7.5,-1.7) [label=right:$C_{bd'^{\square}}$]{};

\node[circle, draw, red, thick, minimum size=0.4cm] (cd0) at (7.5,-2.3) [label=right:$C_{cd^{\square}}$]{};
\node[circle, draw, red, thick,  minimum size=0.4cm] (cd00) at (7.5,-2.9) [label=right:$C_{cd'^{\square}}$]{};

\draw[red, thick] (an') -- (a0);
\draw[red,  thick] (bn') -- (b0);
\draw[red, thick] (cn') -- (c0);
\draw[red, thick] (dn') -- (d0);
\draw[red, thick] (ab_s) -- (ab0);
\draw[red, thick] (ab'_s) -- (ab00);
\draw[red, thick] (ac_s) -- (ac0);
\draw[red, thick] (ac'_s) -- (ac00);
\draw[red, thick] (ad_s) -- (ad0);
\draw[red, thick] (ad'_s) -- (ad00);
\draw[red, thick] (bc_s) -- (bc0);
\draw[red, thick] (bc'_s) -- (bc00);
\draw[red, thick] (bd_s) -- (bd0);
\draw[red, thick] (bd'_s) -- (bd00);
\draw[red, thick] (cd_s) -- (cd0);
\draw[red, thick] (cd'_s) -- (cd00);

\path 
(a) edge (b)
(b) edge (c)
(d) edge (c)
(a) edge (d)
(a) edge (c)
(b) edge (d)
(an) edge (an')
(bn) edge (bn')
(cn) edge (cn')
(dn) edge (dn')
(an) edge (ab)
(an) edge (ac)
(an) edge (ad)
(bn) edge (ab)
(bn) edge (bc)
(bn) edge (bd)
(cn) edge (ac)
(cn) edge (bc)
(cn) edge (cd)
(dn) edge (ad)
(dn) edge (bd)
(dn) edge (cd)
(ab) edge (ab_s)
(ab) edge (ab'_s)
(ac) edge (ac_s)
(ac) edge (ac'_s)
(ad) edge (ad_s)
(ad) edge (ad'_s)
(bc) edge (bc_s)
(bc) edge (bc'_s)
(bd) edge (bd_s)
(bd) edge (bd'_s)
(cd) edge (cd_s)
(cd) edge (cd'_s);
%(ra) edge (an');
\end{tikzpicture}\]
\caption{Reducing {\sc Minimum Maximal Matching}  on planar cubic graphs to 
{\sc Locally Minimal Defensive Alliance} on planar graphs. (a) An undirected graph 
$G=(V,E)$ with minimum maximal matching $M=\{(a,b), (c,d)\}$. (b) The planar graph $G'$ produced by the 
reduction algorithm that has locally minimal defensive alliance 
$ D = A \cup B \setminus \{ab,cd\} \bigcup\limits_{x\in V^{\square} \cup E^{\square}} \Big\{x_i~|~ 1\leq i\leq 60 ~\&~i \mbox{ is not divisible by 3}\Big\} $. A red circle represents a cycle of 
length 60 and a red line between $x$ and $C_x$ indicates that $x$ is adjacent to all the vertices of $C_x$. Note that $G'$ can redrawn in a way in which no edges cross.}
     \label{NP-complete}
 \end{figure}
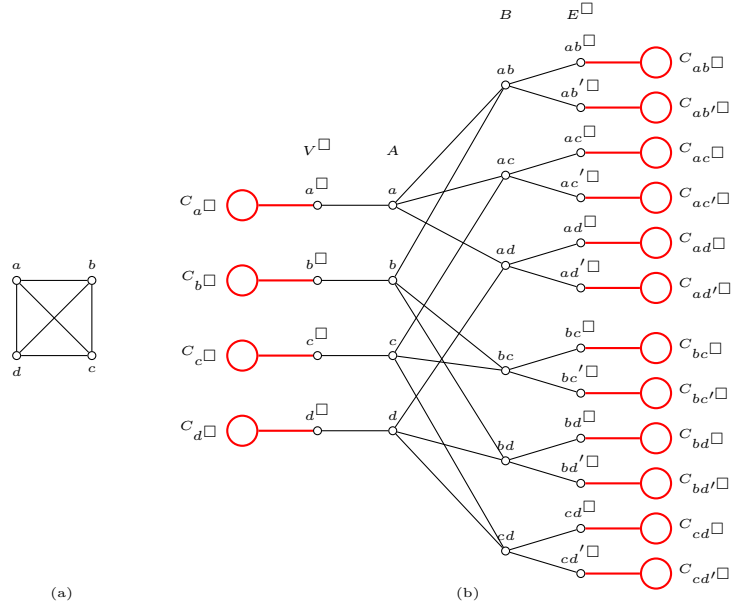

\proof Clearly, the decision version of the problem belongs to NP. 
In order to obtain the NP-hardness result for {\sc Locally Minimal Defensive Alliance}, we obtain a polynomial 
reduction from {\sc Minimum Maximal Matching} on cubic planar graphs proved NP-hard in \cite{MMMHorton}.
Given an instance  $I=(G,k)$ of  {\sc Minimum Maximal Matching}  where $G$ is a cubic planar graph,  
we construct an instance $I'=(G',k')$ of  {\sc Locally Minimal Defensive Alliance} where $G'$ is planar.
See Figure \ref{NP-complete}, which provides an illustration of the construction.
The graph $G^{\prime}$ that we construct has   vertex set $A\cup B$, where
$A=V(G)=\{v_1,v_2,\ldots,v_n\}$, the vertex set of $G$ and 
$B=E(G)=\{e_1,e_2,\ldots,e_m\}$, the edge set of $G$. 
We make $v_i$ adjacent to $e_j$ if and only if $v_i$ is an endpoint of $e_j$. 
Further we add two sets of vertices $V^{\square} = \{v_1^{\square}, \ldots, v_n^{\square}\}$ and 
$E^{\square} = \{e_1^{\square},e_1^{\prime \square}, \ldots,e_m^{\square},e_m^{\prime \square} \}$; and 
make $v_i$ adjacent to $v_i^{\square}$ and $e_i$ adjacent to 
$e_i^{\square}$ and $e_i^{\prime \square}$.
For each  $x\in V^{\square} \cup E^{\square}$, we add a cycle $C_{x}$ of length $6(n+m)$ 
with vertices $x_{1},x_{2},\ldots,x_{6(n+m)}$ and make all the vertices of $C_x$ adjacent to $x$. 
This completes the construction of $G'$. It is easy to note that $G'$ is a planar graph. 
Set $k'= 4(n+m)(n+2m) + (n+m-k)$. To complete the proof, we show that  $G$
has a  maximal matching of size at most $k$ if and only if  $G'$ has a locally minimal
defensive alliance of size at least $k'$.

Suppose $G$ has a maximal matching $M$ of size at most $k$.  
Let $$C^*=\bigcup\limits_{x\in V^{\square} \cup E^{\square}} \Big\{x_i ~|~1\leq i\leq 6(m+n) ~ \& ~i \mbox{ is not divisible by 3}\Big\}.$$ We claim that 
$ D = A \cup (B \setminus M) \cup C^* $ is a locally minimal
defensive alliance of size at least $k'$ in $G'$. Let
$x$ be an arbitrary element of $D$. \\
\noindent{\it Case 1.} Suppose $x$ is an element of $A$. As $G$ is cubic, $x$ has three 
neighbours in $B$. As $M$ is a matching, out of three neighbours of $x$ in $B$,
at most one is in $M$. Therefore, $x$ has at least two neighbours in $B\setminus M$.
Thus, including itself, $x$ has 
at least three defenders and it has at most two attacker. 
Thus every vertex of $A$ is protected. \\
\noindent{\it Case 2.} If $x$ is an element  of $(B \backslash M)$, we will prove that $x$ is marginally protected. The attackers of $x$ consist of its
two neighbours in $E^{\square}$ and the defenders of $x$ consist of its two 
neighbours in $A$. Thus, including itself, $x$ has three defenders and two 
attackers. Hence $x$ is marginally protected. \\
\noindent{\it Case 3.} If $x$ is an element  of  $C^*$, we will prove that $x$ is marginally protected. Without loss of generality 
suppose $x=x_{3i+1}$ for some $i \geq 1$. Then the defender of $x$ is $x_{3i+2}$ and the attackers of $x$ consist of $x_{3i}$ and its only neighbour in 
$V^{\square} \cup E^{\square}$. Therefore, including itself, $x$ has two defenders and 
it has two attackers. Hence $x$ is marginally protected.\\
\noindent This show that $D$ is a defensive alliance. In order to prove that
$D$ is a locally
minimal defensive alliance,  we prove that for any $v\in D$, $D\setminus \{v\}$ is not a defensive alliance. Since $M$ is maximal, it is not
possible to move a vertex from $B\setminus M$  to $D^c$. If we move a vertex from $ B \setminus M$ to
$D^c$, then some vertex from $A \subseteq D$ will become unprotected. 
Similarly it is not possible to move a vertex
from $A$ to $D^c$.  If we move a vertex from $A$ to $D^c$ then some vertices from 
$ B \setminus M\subseteq D$ will become unprotected. As every vertex in $C^*$ has a marginally protected neighbour, we cannot move a vertex from $C^*$ to $D^c$.
This shows that $D$ is indeed a locally minimal defensive alliance.

\par For the reverse direction, suppose  that $G'$  has a locally minimal defensive alliance $D$ of size at least $k'$. Before we continue with the proof for the 
reverse direction, we will prove two crucial properties:
\begin{claim}
Let $D$ be a locally minimal defensive alliance 
in $G'$. Then for each $x\in  V^{\square} \cup E^{\square}$, $C_x$ can contribute at most $4(m+n)$ vertices in $D$. 
\end{claim}

\begin{proof} Let $C_x =\{x_1,\ldots,x_{6(m+n)}\}$. We define the following three sets of vertices $C_{x}^0 = \{x_{3i} ~|~ 1\leq i\leq 2(n+m) \}, C_{x}^1 = \{x_{3i+1} ~|~ 0\leq i\leq  2(n+m)-1 \} $ and $C_{x}^2 = \{x_{3i+2} ~|~ 0\leq i\leq  2(n+m)-1\}$.
 Suppose $x\notin D$. Then the union of any two of the above sets, say, 
 $C_x^1\cup C_x^2$ forms a locally minimal defensive alliance of size $4(m+n)$ 
 and can be part of $D$. 
 The defender of $x_{3i+1}$ is $x_{3i+2}$ and the attackers of $x_{3i+1}$ is $x$ and $x_{3i}$. Thus, including itself, $x_{3i+1}$ has two defenders and two attackers. So $x_{3i+1}$ is marginally protected. Similarly $x_{3i+2}$ is also marginally protected. 
As every vertex of $C_x^1\cup C_x^2$ is protected and  has a marginally protected
neighbour, it is a locally minimal defensive alliance and can be part of $D$. Now 
we prove that $C_x$ cannot contribute more than $4(m+n)$ vertices in $D$. For the 
sake of contradiction, assume that $C_x$ contributes $4(m+n)+1$ vertices in $D$. 
Then $D$ contains at least three consecutive vertices, say, $x_{3i},x_{3i+1},x_{3i+2}$ for some $i$. Again suppose $x\notin D$. Then the defenders of $x_{3i+1}$ consist of 
$x_{3i}, x_{3i+2}$; and the attacker of $x_{3i+1}$ is  $x$ only. Thus, including itself, 
$x_{3i+1}$ has three defenders and one attacker. Thus $x_{3i+1}$ is overprotected. 
It may be verified that $x_{3i}$ and $x_{3i+2}$ are marginally protected but they do not 
have a marginally protected neighbours in $D$. Thus $D$ is not a locally minimal defensive alliance, a contradiction. 
\end{proof}

\begin{claim}
    Let $D$ be a locally minimal defensive alliance of size at least $k'$ in $G'$ and 
    let $x\in V^{\square} \cup E^{\square}$. Then $x$ does not lie in $D$.
    \end{claim}

\begin{proof}
    
For the sake of contradiction assume that  $x\in D$. This means $x$ is either marginally protected or overprotected in $D$.  We consider two cases:\\
\noindent{\it Case 1.} Suppose $x$ is marginally 
protected in $D$. Note that $x$ has $6(m+n)+1$ neighbours in $G'$ and the neighbours of $x$ in $G'$ consists of $6(m+n)$ vertices of $C_x$ and one neighbour in $V^{\square}\cup E^{\square}$. Since $x$ is marginally protected, including itself, it can have $3(m+n)+1$ neighbours in $D$. Thus the cycle $C_{x}$ can contribute at most  $3(n+m)$ vertices in $D$, as otherwise 
$x$ is not marginally protected. Recall that $k'= 4(n+m)(n+2m) + (n+m-k)$ and 
$|V^{\square}\cup E^{\square}|=(n+2m)$. By the above claim we know, for every $y\in V^{\square} \cup E^{\square}$, $C_y$ can contribute at most $4(m+n)$ vertices in $D$.
Assume that for each $y\in V^{\square} \cup E^{\square}$ $(y\neq x)$,  $y$ does not belong to $D$ and $C_y$ contributes exactly $4(m+n)$ vertices in $D$.  On the other hand $x$ is marginally protected in $D$ and $C_x$ contributes
at most $3(m+n)$ vertices in $D$. Even if we include all the vertices of
$(A\cup B) \setminus \{x\}$ in $D$, its size  reaches the value  
$4(m+n)(n+2m-1)+ 3(m+n)+(m+n-1) <k'$ and this is a contradiction. Thus $x$ does not 
lie in $D$.\\
\noindent{\it Case 2.} Suppose $x$ lies in $D$ and it is overprotected. Let $x_i$ be a neighbour of $x$ in $C_x \cap D$. 
As $D$ is a locally minimal defensive alliance, $x_i$ must have a  marginally protected neighbour, say 
$x_{i+1}$, in $C_x$. The defenders of $x_i$ consist of $x_{i+1}$ and $x$. Thus, including itself, the number
of defenders of $x_i$ is at least 3, and the number of attackers is at most 1. 
Hence $x_i$ is not marginally protected;  similarly $x_{i+1}$ is also not marginally protected. 
This implies that $x_{i}$ and $x_{i+1}$ do not have a marginally protected neighbour,
a contradiction to the assumption that $D$ is a locally minimal defensive alliance.
Thus $x$ does not lie in $D$.
This proves the claim that  $(V^{\square}\cup E^{\square})\cap D=\emptyset $. 
\end{proof}
%As $x\in V^{\square} \cup E^{\square}$ is not in $D$, then $C_{x}$ can contribute at most $4(n+m)$ vertices in $D$ by choosing two sets from the following three sets of vertices $X_{0} = \{x_{3i} ~|~ 1\leq i\leq 2(n+m) \}, X_{1} = \{x_{3i+1} ~|~ 0\leq i\leq  2(n+m)-1 \} $ and $X_{2} = \{x_{3i+2} ~|~ 0\leq i\leq  2(n+m)-1\}$.
We now define $M=B\cap D^c$ and claim that there exists a set $M'\subseteq M$ such that  $M'$ is a maximal matching. Assume, for the sake of contradiction, that no subsets of $M$
form a maximal matching. 
If no subsets of $M$ form a maximal matching then clearly there exists an edge $(u,v)\in E$ 
such that it can still be added to $M$. This means no edges incident with $u$ and $v$
are  in $M$. 
It implies that all the edges incident with $u$ and $v$ are in $D$ and 
hence $u$ and $v$ are overprotected. Note that vertex $e=uv\in B$ has four neighbours
$u,v,uv^{\square}$ and $uv'^{\square}$. By the above claim $uv^{\square}$ and $uv'^{\square}$ are not in $D$. The other two neighbours $u$ and $v$ are overprotected in $D$.
Therefore, vertex $e=uv \in B$ does not have a marginally protected neighbour in $D$ 
as both of its neighbours $u$ and $v$ are overprotected, a 
contradiction to the fact that $D$ is a locally minimal defensive alliance. 
This shows that there must exist a set $M'\subseteq M$ such that $M'$ is a 
maximal matching and $|M'|\leq |M|\leq k$. This completes the proof of Theorem 
\ref{Ffvsthm}. \qed

\section{A randomized FPT algorithm for {\sc Exact Connected Locally Minimal Defensive Alliance} parameterized by solution size}

In this section, we give a randomized FPT algorithm for  {\sc Exact Connected 
Locally Minimal Defensive Alliance} parameterized by the solution size $k$.
Let $G=(V,E)$ be a graph and let $S\subseteq V$ be a subset of size $k$. 
Every vertex in $G$
 is colored independently with one colour from the set $\{\mbox{red}, \mbox{green}\}$ with uniform 
 probability. Denote the obtained coloring by  $\chi~ : ~V(G) \rightarrow \{\text{red, green}\}$.
 A connected locally minimal  defensive alliance $S$ in $G$ is called a 
 \emph{green connected locally minimal 
 defensive alliance} in $G$ with coloring $\chi$ if all the vertices in $S$ are colored with green color and all the vertices 
 in $N(S)$ are colored red.

\begin{lemma}\label{colorlemma1} 
Let $G$ be a graph and let $\chi : V(G) \rightarrow \{\text{red, green}\}$ be a colouring  of its vertices
 with two colours, chosen uniformly at random. Let $S\subseteq V$ be a connected
 locally minimal  defensive alliance of size $k$ in $G$.  
 Then the probability that
 the elements of 
 $S$ are coloured with green colour and elements of 
 $N(S)$ are coloured with red colour is at least $\frac{1}{2^{k^2+k}}$.
\end{lemma}

\proof  As $S$ is a (connected locally minimal) defensive alliance of size 
$k$ in $G$, each element $v$ in $S$ is protected and therefore 
$v$ can have at most $k$ neighbours outside $S$. It follows that
$|N(S)|\leq |S|k =k^2$. 
 There are $2^n$ possible colorings $\chi$; and there are 
$2^{n-k^2-k}$ possible colorings where the $k$ vertices of $S$ are 
coloured green and at most $k^2$ neighbours of $S$ are colored red. 
Hence the lemma follows. \qed
 
 \begin{lemma}\label{colorlemma2}
 Let $G$ be a graph and let $\chi : V(G) \rightarrow \{\text{red, green}\}$ be a colouring  of its vertices
 with two colours.  Then there exists an algorithm that checks in time $O(n+m)$ 
 whether $G$ contains a green connected locally minimal 
 defensive alliance of size $k$ and, if this is the case, returns one such an alliance.
 \end{lemma}
 
 \proof Let $V_g$ and $V_r$ be a partitioning of $V(G)$ such that all vertices in 
 $V_g$ are coloured green and all vertices in $V_r$ are coloured red. A connected
 component $C$ is said to be a green connected component if the vertices of
 $C$ are colored green.  Run DFS to identify 
 all green connected
 components $C_1,C_2,\ldots,C_{\ell}$ of $G[V_g]$ in $O(m+n)$ time.  Then verify in linear time
 if there exists a green connected component $C_i$ of size 
 $k$ that forms a locally minimal defensive alliance in $G$.  \qed
 \vspace{10pt}
\noindent We now combine Lemma \ref{colorlemma1} and Lemma \ref{colorlemma2} to obtain 
the main result of this section.

\begin{theorem}\label{ETH} 
There exists a randomized algorithm that, given an {\sc Exact Connected Locally Minimal 
Defensive Alliance} 
instance $(G, k)$, in time ${2^{\mathcal{O}{(k^{2}+k)}}(n+m)}$ either reports a failure or 
finds a  connected locally  minimal 
defensive alliance of size exactly $k$ in $G$. Moreover, if the algorithm is given a yes-instance, it returns a solution with a constant probability.
 \end{theorem}
 
\proof Given an input instance $(G, k)$, we uniformly at random color the vertices of $V(G)$ with two colors green and red. That is, every vertex is colored independently with either green or red color with uniform probability.  
Denote the obtained coloring by $\chi : V(G) \rightarrow \{\text{red, green}\}$. 
We run the algorithm of Lemma \ref{colorlemma2} on the graph $G$ with coloring $\chi$.
If it returns a green connected locally minimal  defensive alliance  $S$ of size $k$, 
then we return
this $S$ as connected locally minimal  defensive alliance of size $k$ in $G$. Otherwise, we report
failure.

It remains to bound the probability of finding a connected locally minimal  defensive alliance 
of size $k$ in the case $(G,k)$ is a yes-instance. To this end, suppose $G$ has a 
connected locally minimal  defensive alliance $S$ of size $k$. By Lemma \ref{colorlemma1},
$S$ becomes a green connected locally minimal defensive alliance of size $k$ in the colouring
$\chi$ with probability at least $\frac{1}{2^{k^2+k}}$. If this is the case, the algorithm of 
Lemma \ref{colorlemma2} finds a green connected locally minimal  defensive alliance of size $k$
(not necessarily $S$ itself), and the algorithm returns a connected locally minimal 
defensive
alliance of size $k$ in $G$. 

Thus we have an algorithm that runs in time $O(m+n)$ and given a yes-instance, returns 
a solution with probability at least $\frac{1}{2^{k^2+k}}$. Clearly, by repeating the algorithm
independently $2^{k^2+k}$ times, we obtain the running time bound and the success probability
at least $1-\frac{1}{e}$. \qed

\section{FPT algorithm parameterized by neighbourhood diversity}\label{ndsection}

In this section, we present an FPT algorithm for {\sc Locally Minimal Defensive Alliance}  parameterized 
 by neighbourhood diversity. That is, we prove the following theorem.
\begin{theorem}\label{theoremnd1}
        {\sc Locally Minimal  Defensive Alliance}  is fixed-parameter tractable when parameterized by  the neighbourhood diversity.
    \end{theorem}
 
 In a graph $G=(V,E)$, we say two vertices $u$ and $v$ have the same \emph{type} if and only if 
 $N(u)\setminus \{v\}=N(v)\setminus \{u\}$. The relation of having the same type 
 is an equivalence  relation. The idea of neighbourhood diversity is based on this 
 type structure. 
  \begin{definition} \rm \cite{Lampis} 
        The neighbourhood diversity of a graph $G=(V,E)$, denoted by ${\tt nd}(G)$, is the least integer $k$ for which we can partition the set $V$ of vertices  into $k$ classes, such that all vertices in each class have the 
        same type.
    \end{definition}
    
\begin{figure}[ht]
\centering
\begin{tikzpicture}[scale=0.8]
\node[fill=pink,circle,draw, minimum size=0.1cm] (a) at (2, 0) [label=above:$a$]{};
\node[fill=pink,circle,draw, minimum size=0.1cm] (b) at (3, 0) [label=above:$b$]{};
\node[fill=pink,circle,draw, minimum size=0.1cm] (c) at (4, 0) [label=above:$c$]{};
\node[fill=pink,circle,draw, minimum size=0.1cm] (d) at (5, 0) [label=above:$d$]{};
\node[fill=white,circle,draw, minimum size=0.1cm] (e) at (3.5, -2) [label=right:$e$]{};
\node[fill=yellow,circle,draw, minimum size=0.1cm] (f) at (2.5, -4) [label=below :$f$]{}; 
\node[fill=yellow,circle,draw, minimum size=0.1cm] (g) at (1.5, -3) [label=below :$g$]{};
\node[fill=babyblue,circle,draw, minimum size=0.1cm] (h) at (5.5, -3) [label=below :$h$]{};
\node[fill=babyblue,circle,draw, minimum size=0.1cm] (i) at (4.5, -4) [label=below :$i$]{};
\node (g1) at (3.5, -5) [label=below :$G$]{}; 
\node (h1) at (11, -5) [label=below :$H$]{}; 

\path

(a) edge (e)
(b) edge (e)
(c) edge (e)
(d) edge (e)
(g) edge (e)
(f) edge (e)
(h) edge (e)
(i) edge (e)
(g) edge (f)
(g) edge (h)
(g) edge (i)
(f) edge (h)
(f) edge (i);

\node[fill=pink, circle,draw, minimum size=1cm ] (a0) at (11,0) []{{$a,b,c,d$}};
\node[fill=white, circle,draw, minimum size=1cm ] (b0) at (11,-2) []{{$e$}};	
\node[fill=yellow, circle,draw, minimum size=1cm ] (c0) at (9,-4) []{{$f,g$}};
\node[fill=babyblue, circle,draw, minimum size=1cm ] (d0) at (13,-4) []{{$h,i$}};	

\path 
(a0) edge (b0)
(b0) edge (c0)
(b0) edge (d0)
(c0) edge (d0);

\end{tikzpicture}
\caption{A graph $G$ with neighbourhood diversity 4 and its corresponding type graph $H$. }
\label{ndfig}
\end{figure}
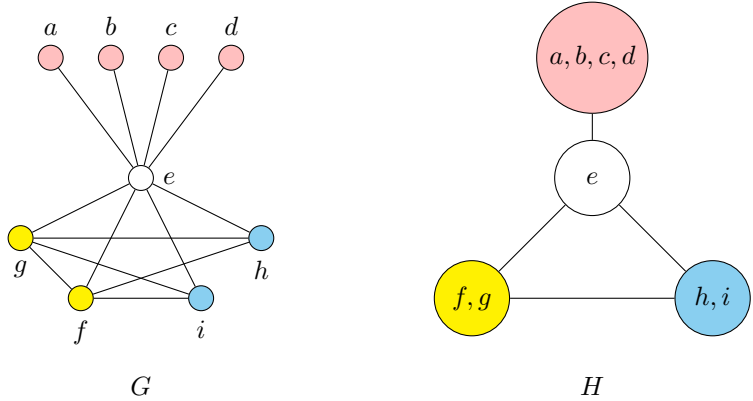

    If neighbourhood diversity of a graph is bounded by an integer $k$, then there exists 
    a partition $\{ T_1, T_2,\ldots, T_k\}$ of $V(G)$ into $k$ type  classes. It is known that 
    such a minimum partition can be found in linear time using fast modular decomposition algorithms \cite{Tedder}. 
    %It is also known that ${\tt nd}(G) \leq 2^{{\tt vc}(G)}+{\tt vc}(G)$ for every graph $G$, where
    %{\tt vc}(G) denotes the size of a minimum vertex cover of $G$ \cite{Lampis}. 
    Notice
    that each type class  could either be a clique or an independent set by definition.  
     For algorithmic 
    purpose it is often useful to consider a {\it type graph} $H$ of  graph $G$, where
    each vertex of $H$ is a type class in $G$, and two vertices $T_i$ and $T_j$ are adjacent if and only if
    there is complete bipartite clique between these type classes in $G$. It is not difficult to see that
    there will be either a complete bipartite clique or no edges between any two type classes. 
      The key property of graphs of
  bounded neighbourhood diversity is that their type graphs have bounded size.
  For example, a graph $G$ with neighbourhood diversity four and its corresponding type graph $H$ is illustrated in Figure \ref{ndfig}.
  \\

 \noindent {\it Outline of the algorithm.} Given an $n$-vertex graph $G$ with ${\tt nd}(G) \leq k$, 
 we find a partition $\{ T_1, T_2,\ldots, T_k\}$ of $V(G)$ into $k$ type  classes. It is known that 
    such a minimum partition can be found in linear time using fast modular decomposition algorithms \cite{Tedder}. Suppose $S$ is a hypothetical  locally 
    minimal defensive alliance in $G$. Next we guess whether a type class $T_i$ contributes no vertices, one vertex or at least two vertices to $S$.  There are at most 
    $3^k$ guesses as each $T_i$ has three options: either it contributes no vertex, one vertex, or at least two vertices to $S$.  Finally we reduce the problem of finding  $S$ to an integer linear programming (ILP) optimization with at most $k$ variables. Since ILP optimization is fixed-parameter tractable when parameterized by the number of variables \cite{fellows}, we can conclude that our problem is fixed-parameter tractable when parameterized by the neighbourhood diversity.  

     \subsection{Characterization of a locally minimal defensive alliance $S$ with 
     type classes}
     Let $G$ be a connected graph such that ${\tt nd}(G)=k$. In this section we assume that we have  the partition 
    of $V(G)$ into sets of type classes $T_1,\ldots,T_k$. We assume $k\geq 2$ since otherwise the problem 
     becomes trivial. We prove the following lemma.
     \begin{lemma}\label{S1S2}
        Suppose $S_1,S_2\subseteq V(G)$ are such that $|S_1\cap T_i|=|S_2\cap T_i|$ for all
        $i\in [k]$. Then $S_1$ is a locally minimal defensive alliance in $G$ if and only if  $S_2$ is also a locally minimal defensive alliance in $G$. 
        \end{lemma}
        \proof Suppose $S_1$ is a locally minimal defensive alliance in $G$. For each $i\in [k]$,
        the vertices in $S_1\cap T_i$ and $S_2\cap T_i$  have the same neighbourhood
        in $G$ 
        as the vertices in $T_i$ have the same neighbourhood in $G$.
        Therefore $S_2$ is also a locally minimal defensive alliance in $G$.
        The converse part of the lemma also holds. This completes the proof of Lemma \ref{S1S2}. \qed 
     
     \par Suppose $S$ is a hypothetical locally minimal defensive alliance. Let $\mathcal{T}_0$ (resp. $\mathcal{T}_1$) be the set of all type classes  that contribute
     zero vertices (one vertex) to $S$. Similarly, let $\mathcal{T}_{\geq 2}$ be the set
     of all type classes that contribute at least two vertices to $S$.
     More formally, we define the following sets:
     $$\mathcal{T}_0=\{ T_i~:~|S\cap T_i|=0\};~ \mathcal{T}_{1}=\{ T_i~:~|S\cap T_i|= 1\};~ \mathcal{T}_{\geq 2}=\{ T_i~:~|S\cap T_i|\geq 2\}.$$
Given $\mathcal{T}_0$, $\mathcal{T}_{1}$ and $\mathcal{T}_{\geq 2}$, our goal here is to find a largest locally
        minimal defensive alliance $S$ of  $G$, with $|S\cap T_i|=0$ when $T_i\in \mathcal{T}_0$,
        $|S\cap T_i|=1$ when $T_i \in  {\mathcal{T}_{1}}$ and $|T_i\cap S|\geq 2$ when $T_i \in  {\mathcal{T}_{\geq 2}}$.
             Let $x_i=|S\cap T_i|$ for $i\in [k]$. We partition $[k]$ into $I_0, I_1$ and $I_{\geq 2}$ as follows:
     $$i\in  \begin{cases}
  I_0  &  x_i=0  \\
  I_1 & x_i=1\\
  I_{\geq 2} & x_i\geq 2
\end{cases}$$        
By Lemma \ref{S1S2},
        the variables $x_i$ determine
        $S$ uniquely.  
        Therefore, given $I_0,I_1,I_{\geq 2}$, the goal   is to 
        maximize the sum $$ \sum\limits_{i\in[k]}{1\cdot [T_i\in \mathcal{T}_1]}+  \sum\limits_{i\in[k]}{x_i \cdot [T_i\in \mathcal{T}_{\geq 2}]}=|I_1|+ 
        \sum\limits_{i\in I_{\geq 2}}{x_i}$$ under the conditions: $x_i=0$ for
        $i\in I_0$, $x_i=1$ for $i\in I_1$,
        $2\leq x_i\leq |T_i|=n_i$  for $i\in I_{\geq 2}$ and two additional conditions
        (Type 1 and Type 2) described below. Here $[T_i\in \mathcal{T}_1]$ and $[T_i\in \mathcal{T}_{\geq 2}]$  are the Iverson brackets; see Definition \ref{Iverson}. \\
        
\noindent{\bf Type 1 Condition:}   For each $i\in I_1 \cup I_{\geq 2}$, we add the condition given in Equation 1. This is called type 1 condition. Type 1 conditions ensure  that $S$ is a 
defensive alliance, that is, the vertices in $T_i \cap S$ are protected 
for all $i\in I_1\cup I_{\geq 2}$. Define 
$$\mathcal{K}=\text{the collection of all 
clique type classes.} $$  
%For each $v\in C_j$ the number of neighbours of $v$ in $S$ is given by 
%\begin{align}\label{eq1}
%  $$(x_j-1)[C_j\in \mathcal{K}]+ \sum\limits_{i\in [k]}{1\times  [C_i\in N_H(C_j) \cap \mathcal{C}_1]}+  \sum\limits_{i\in [k]}{x_i [C_i\in N_H(C_j) \cap \mathcal{C}_{\geq 2}]}$$
%\end{align}
%         Thus, including itself, $v$ has $1+(x_j-1)[C_j\in \mathcal{C}]+ \sum\limits_{i\in [k]}{x_i [C_i\in N_H(C_j) \cap I_1]}$ defenders in  $G$. Note that if $C_i\in I_1$, then only $x_i$  vertices of $C_i$ are in $S$ and the the remaining $n_i-x_i$ vertices of  $C_i$ are outside $S$.  The number of neighbours of $v$ outside $S$ is given by   $$ (n_j-x_j)[C_j\in \mathcal{C}]+ \sum\limits_{i\in [k]}{(n_i-x_i)[C_i\in N_H(C_j) \cap {I_1}]} + \sum\limits_{i\in[k] }{n_i~[C_i\in N_H(C_j)\cap I_0]}$$
A vertex $u\in T_i\cap S$ 
is protected if and only if    $d_S(u)\geq \frac{d_G(u)-1}{2} $, that is,
\begin{align}\label{eq2+}
&(x_i-1)[T_i\in \mathcal{K}]+ \sum\limits_{j\in [k]}{1\cdot  [T_j\in N_H(T_i) \cap \mathcal{T}_1]} + \sum\limits_{j\in [k]}{x_j \cdot [T_j\in N_H(T_i) \cap \mathcal{T}_{\geq 2}]} \nonumber \\
&\geq \frac{d_G(u)-1}{2}
 \end{align}
 The left-hand side expression of Eq. \ref{eq2+} is equal to $d_S(u)$. The first term of the 
 expression stands for the number of neighbours of $u$ in $S\cap T_i$ if $T_i$
 is a clique type class; the second term  stands for the number 
 of neighbours of $u$ in $S\cap T_j$ if $T_j\in \mathcal{T}_1$ and $T_j$ is a neighbour 
 of $T_i$ in the type graph $H$; the third term  stands for the number 
 of neighbours of $u$ in $S\cap T_j$ if $T_j\in \mathcal{T}_{\geq 2}$ and $T_j$ is a neighbour 
 of $T_i$ in the type graph $H$.\\
 \noindent{\bf Type 2 Condition:}         
%Let ${\bf x}=(x_1,\ldots,x_k)$ be the vector corresponds to  $S\subseteq V(G)$. 
%We want to make sure that the vector ${\bf x}=(x_1,\ldots,x_k)$ or  $S$ forms a locally minimal defensive alliance, that is, for any $v\in S$, $S\setminus \{v\}$ is not a defensive alliance.
 Type 2 condition ensures  that  $S$ is a locally minimal defensive 
 alliance, that is, for any $v\in S$, $S\setminus \{v\}$ is not a defensive alliance. 

 \begin{lemma} \label{charac-LMDA} Suppose $S$ is a defensive alliance in $G$.
 Then, given $I_0,I_1,I_{\geq 2}$, $S$ is a locally minimal defensive alliance in $G$
   if and only if 
there is a function $f: I_1 \cup I_{\geq 2} \rightarrow I_1\cup I_{\geq 2}$ 
with $f(i)\neq i$ when $i\in I_1$
such that  for each $i$ the vertices of  $(S \setminus \{v\})\cap T_{f(i)}$ are
    unprotected in $S\setminus \{v\}$ for any $v\in S\cap T_i$.
\end{lemma} 
\begin{proof} The proof follows directly from the definition of locally minimal defensive alliance. 
%We assume that we have  the partition  of $V(G)$ into sets of type classes $T_1,\ldots,T_k$. Then $$S=\bigcup_{i\in I_1\cup I_{\geq 2}} S\cap T_i. $$
Suppose $S$ is a locally minimal defensive alliance in $G$. 
    We know for any $v\in S$, $S\setminus \{v\}$ is not a defensive alliance, that is, 
    some vertices of $S\setminus \{v\}$ are unprotected. 
    In terms of type classes, we can say there exists a function $f: I_1 \cup I_{\geq 2} \rightarrow I_1\cup I_{\geq 2}$ 
with $f(i)\neq i$ when $i\in I_1$
such that for each $i$ the  vertices of $(S \setminus \{v\})\cap T_{f(i)}$ are 
    unprotected in $S\setminus \{v\}$ for any $v\in S\cap T_i$.
    \par To prove the reverse direction, suppose $S$ is a defensive alliance and 
    there is a function $f: I_1 \cup I_{\geq 2} \rightarrow I_1\cup I_{\geq 2}$ 
with $f(i)\neq i$ when $i\in I_1$
such that  for each $i$ the vertices of  $(S \setminus \{v\})\cap T_{f(i)}$ are
    unprotected in $S\setminus \{v\}$ for any $v\in S\cap T_i$.
 This implies that for each $i$, every vertex of $S\cap T_i$ has a marginally protected
    neighbour in $ S\cap T_{f(i)}$. Thus 
    $S=\bigcup\limits_{i\in I_1\cup I_{\geq 2}} S\cap T_i$ is a locally minimal defensive alliance. This completes the proof of Lemma \ref{charac-LMDA}.
        \end{proof}
\noindent Given $(I_0,I_1,I_{\geq 2})$ and $f$, we want $S$ to satisfy the conditions in Equation 2. 
This is called type 2 condition. Type 2 conditions ensure that $S$ is a locally
minimal defensive alliance.  By Lemma 
\ref{charac-LMDA}, the vertices in $ (S\setminus \{v\})\cap T_{f(i)}$ 
must be unprotected in $S\setminus \{v\}$ for any $v\in T_{i}$. A vertex $u$ from  
$ (S\setminus \{v\})\cap T_{f(i)}$ is unprotected in $S\setminus \{v\}$  if and only if
 the number of neighbours of $u$ in $S\setminus \{v\}$ is strictly less than $ \frac{d_G(u)-1}{2}$, that is, 
\begin{align}\label{eq2++}
&(x_{f(i)}-1)[T_{f(i)}\in \mathcal{K}]+ (x_{i}-1)[T_{i}\in N_H(T_{f(i)})] + \sum\limits_{j\in [k]; j\neq i}{1\cdot  [T_j\in N_H(T_{f(i)}) \cap \mathcal{T}_1]}+ \nonumber\\
 & \sum\limits_{j\in [k]; j\neq i}{x_j\cdot [T_j\in N_H(T_{f(i)}) \cap \mathcal{T}_{\geq 2}]}
< \frac{d_G(u)-1}{2}
\end{align} 
The left-hand side expression of Eq. \ref{eq2++} is equal to $d_{S\setminus \{v\}}(u)$ where 
$u\in (S \setminus \{v\})\cap T_{f(i)}$. The first term of the 
 expression stands for the number of neighbours of $u$ in $S\cap T_{f(i)}$ if $T_{f(i)}$
 is a clique type class; the second term  stands for the number 
 of neighbours of $u$ in $(S \setminus \{v\})\cap T_i$ if  $T_i$ is a neighbour 
 of $T_{f(i)}$ in the type graph $H$; 
 the third term  stands for the number 
 of neighbours of $u$ in $S\cap T_j$ if $T_j\in \mathcal{T}_1$ and $T_j$ is a neighbour 
 of $T_{f(i)}$ in the type graph $H$; the fourth term  stands for the number 
 of neighbours of $u$ in $S\cap T_j$ if $T_j\in \mathcal{T}_{\geq 2}$ and $T_j$ is a neighbour 
 of $T_{f(i)}$ in the type graph $H$.

 \subsection{ILP formation for {\sc Annotated LMDA}}
Our algorithm for {\sc Locally Minimal Defensive Alliance (LMDA)} will use the following annotated problem as 
subroutine. In the {\sc Annotated LMDA} problem, we are given a graph 
$G$ with ${\tt nd}(G)=k$, type classes $T_1,T_2,\ldots,T_k$ of $G$, a partition of $[k]$ into three parts  $I_0, I_1,I_{\geq 2}$, a function
$f: I_1\cup I_{\geq 2}\rightarrow I_1\cup I_{\geq 2} $ with $f(i)\neq i$ if $i\in I_1$and the goal is to find a
largest locally minimal defensive alliance  
$S\subseteq V(G)$ 
such that $|S\cap T_i|=0$ if $i\in I_0$, $|S\cap T_i|=1$ if $i\in I_1$ and 
$|S\cap T_i|\geq 2$ if $i\in I_{\geq 2}$.

 Let $(G, (T_1,\ldots,T_k), I_0, I_1, I_{\geq 2},~ f)$ be an instance of 
{\sc Annotated LMDA}.  We reduce the problem of solving {\sc Annotated LMDA}
    to an integer linear programming  optimization with at most $k$ variables as follows:\\
    
\noindent       
\vspace{4mm}
    \fbox
    {\begin{minipage}{33.7em}\label{Min-FFVS}      
\begin{equation*}
\begin{split}
&\text{Maximize }  |I_1|+ \sum\limits_{i\in I_{\geq 2}}{x_i} \\
%&\text{Maximize} \sum\limits_{C_i\in I_1\cup I_2}{x_i} \\
&\text{Subject to~~~} \\
&  x_i=0  ~~\text{for}~~ i\in I_0\\ 
&  x_i=1  ~~\text{for}~~ i\in I_1\\ 
&  2\leq x_i \leq |C_i|=n_i ~~ \text{for}~~i\in I_{\geq 2} \\ % &  x_{i}=1 \text{ for all }i ~:~ C_i\in  I_1;\\
%&                                         x_{i} \in \{2,\ldots, |C_i| \} \text{ for all }i ~:~ C_i\in  I_2\\
& \text {Equation (1)} ~~\text{for} ~~ i\in I_1\cup I_{\geq 2} \\
& \text {Equation (2)} ~~\text{for} ~~ i \in I_1\cup I_{\geq 2}\\
%&\text {Equation (3)} ~~\forall ~i_l\in I_1 ~~\forall ~j\in [r-1]  \\
%&\text {Equation (3)} ~~\forall ~i_l\in I_{\geq 2} ~~\forall~ j\in[r]  \\
\end{split}
\end{equation*}  
  \end{minipage} }    
  
\subsection{Running time for {\sc Annotated LMDA}}
Lenstra \cite{lenstra} showed that the feasibility version of {\sc $k$-ILP} is FPT with 
running time doubly exponential in $k$, where $k$ is the number of variables. 
Later, Kannan \cite{kannan} proved an algorithm for {\sc $k$-ILP} running in time $k^{O(k)}$.
In our algorithm, we need the optimization version of {\sc $k$-ILP} rather than 
the feasibility version. We state the minimization version of {\sc $k$-ILP}
as presented by Fellows et al. \cite{fellows}. \\

\noindent {\sc $k$-Variable Integer Linear Programming Optimization ($k$-Opt-ILP)}: Let matrices $A\in \ Z^{m\times k}$, $b\in \ Z^{k\times 1}$ and 
$c\in \ Z^{1\times k}$ be given. We want to find a vector $ x\in \ Z ^{k\times 1}$ that minimizes the objective function $c\cdot x$ and satisfies the $m$ 
inequalities, that is, $A\cdot x\geq b$.  The number of variables $k$ is the parameter. 
Then they showed the following:

\begin{lemma} \label{ilp}\cite{fellows} 
 {\sc $k$-Opt-ILP} can be solved using $O(k^{2.5k+o(k)}\cdot L \cdot \log(MN))$ arithmetic operations and space polynomial in $L$. 
Here $L$ is the number of bits in the input, $N$ is the maximum absolute value any variable can take, and $M$ is an upper bound on 
the absolute value of the minimum taken by the objective function.
\end{lemma}

In the formulation for {\sc Annotated LMDA}, we have at most $k$ variables. The value of objective function is bounded by $n$ and the value of any variable 
in the integer linear programming is also bounded by $n$. The constraints can be represented using 
$O(k^2 \log{n})$ bits. Lemma \ref{ilp} implies that we can solve the problem   in FPT time. 

\subsection{An algorithm for {\sc Locally Minimal Defensive Alliance (LMDA)}}

\begin{lemma} \label{annotated to GMDA}
If there exists an FPT algorithm for {\sc Annotated LMDA} then there exists an 
FPT algorithm for {\sc LMDA}.
\end{lemma}

\proof Suppose there exists an FPT algorithm for {\sc Annotated LMDA} parameterized 
by neighbourhood diversity $k$. Note that there are $3^k k^k$ candidates for {\sc Annotated LMDA} instances. 
The reason is this. There are at most $3^k$ candidates for $(I_0,I_1, I_{\geq 2})$  as each $T_i$ has three options: either in $I_0$, $I_{1}$ or $I_{\geq 2}$; 
there are at most $k^k$ candidates for 
$f$.  In order to obtain a  locally minimal 
defensive alliance of maximum size, we first solve all {\sc Annotated LMDA} instances, then consider 
a largest solution over all {\sc Annotated LMDA} instances.   Therefore,  
{\sc LMDA} can be solved in FPT time parameterized by neighbourhood diversity $k$. \qed
\vspace{10pt}

\noindent We  proved that the ILP formula for an {\sc Annotated GMDA}  can be solved in FPT time. By Lemma \ref{annotated to GMDA}, thus
Theorem \ref{theoremnd1} holds.

\section{Hardness  of {\sc Locally Minimal Defensive Alliance} parameterized by treewidth}
\label{W[1]-section}
In this section we show that 
{\sc Locally Minimal Defensive Alliance}
is W[1]-hard parameterized by treewidth, via a reduction 
from a variant of {\sc Defensive Alliance}. The input of {\sc Defensive Alliance} consists 
of a graph $G$, and an integer $k$, the task is to decide if $G$ has  a defensive 
alliance of size at most $k$.  Gaikwad and Maity \cite{GAIKWAD2022136} proved   that {\sc Defensive Alliance}  is W[1]-hard when parameterized by  
the size of a vertex deletion set into collection of stars, i.e., the size
of a subset $D$ of the vertices of the graph such that every component in the graph, after removing $D$, is a 
star.  They proved the following result.
\begin{theorem}\cite{GAIKWAD2022136}\label{FNvds}
 {\sc Defensive Alliance} is W[1]-hard when parameterized by  the size of a vertex deletion set into collection of stars, even when restricted to bipartite graphs.
 \end{theorem}
\noindent  While {\sc Defensive Alliance}  asks for a defensive alliances of size at most $k$, we  consider  {\sc Exact  Defensive Alliance}  that concerns defensive alliances of size exactly $k$.  
\noindent As a consequence of Theorem \ref{FNvds},  we have the following result:  
 \begin{corollary}\label{FNvdsExact}
 {\sc Exact Defensive Alliance} is W[1]-hard when parameterized by  the size of a vertex deletion set into collection of stars, even when restricted to bipartite graphs.
 \end{corollary}
  In this section, we prove the following theorem:
 \begin{theorem}\label{twtheorem}
 {\sc Locally Minimal Defensive Alliance} is W[1]-hard when parameterized by  the size of a vertex deletion set into collection of stars. 
 %, even when restricted to bipartite graphs.
 \end{theorem}

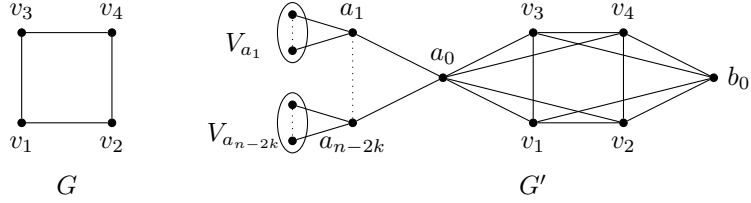
\begin{figure}[ht]
 \begin{tikzpicture}[scale=0.8]
     \node[circle,fill=black, draw, inner sep=0 pt, minimum size=0.1cm](v1) at (0,0) [label=below:$v_{1}$]{};
    \node[circle,fill=black, draw, inner sep=0 pt, minimum size=0.1cm](v2) at (1.5,0) [label=below:$v_{2}$]{};
    \node[circle,fill=black, draw, inner sep=0 pt, minimum size=0.1cm](v3) at (0,1.5) [label=above:$v_{3}$]{};
    \node[circle,fill=black, draw, inner sep=0 pt, minimum size=0.1cm](v4) at (1.5,1.5) [label=above:$v_{4}$]{};

 \node[circle,fill=black, draw, inner sep=0 pt, minimum size=0.1cm](b0) at (3,0.75) [label=right:$b_{0}$]{};

  \node[circle,fill=black, draw, inner sep=0 pt, minimum size=0.1cm](a0) at (-1.5,0.75) [label=above:$a_{0}$]{};

  \node[circle,fill=black, draw, inner sep=0 pt, minimum size=0.1cm](a1) at (-3,1.5) [label=above:$a_{1}$]{};
  \node[circle,fill=black, draw, inner sep=0 pt, minimum size=0.1cm](a2) at (-3,0) [label=below:$a_{n-2k}$]{};

\draw (-4,1.5) ellipse (.25cm and .5cm);
\draw (-4,0) ellipse (.25cm and .5cm);

  \node[circle,fill=black, draw, inner sep=0 pt, minimum size=0.1cm](x1) at (-4,-0.3) []{};
\node[circle,fill=black, draw, inner sep=0 pt, minimum size=0.1cm](x2) at (-4,0.3) []{};
 \node[circle,fill=black, draw, inner sep=0 pt, minimum size=0.1cm](x3) at (-4,1.8) []{};
  \node[circle,fill=black, draw, inner sep=0 pt, minimum size=0.1cm](x4) at (-4,1.2) []{};

\node (va1) at (-4.8,1.8) [label=below:$V_{a_{1}}$]{};
\node (va1) at (-4.8,.3) [label=below:$V_{a_{n-2k}}$]{};

\draw(v1)--(v2);
\draw(v1)--(v3);
\draw(v3)--(v4);
\draw(v4)--(v2);
\draw(a0)--(v1);
\draw(a0)--(v2);
\draw(a0)--(v3);
\draw(a0)--(v4);
\draw(b0)--(v1);
\draw(b0)--(v2);
\draw(b0)--(v3);
\draw(b0)--(v4);
\draw(a0)--(a1);
\draw(a0)--(a2);
\draw[dotted](a2)--(a1);
\draw(a1)--(x3);
\draw(a1)--(x4);
\draw(a2)--(x1);
\draw(a2)--(x2);
\draw[dotted](x2)--(x1);
\draw[dotted](x4)--(x3);

\node[circle,fill=black, draw, inner sep=0 pt, minimum size=0.1cm](v1) at (-7,0) [label=below:$v_{2}$]{};
\node[circle,fill=black, draw, inner sep=0 pt, minimum size=0.1cm](v2) at (-8.5,0) [label=below:$v_{1}$]{};
\node[circle,fill=black, draw, inner sep=0 pt, minimum size=0.1cm](v3) at (-7,1.5) [label=above:$v_{4}$]{};
\node[circle,fill=black, draw, inner sep=0 pt, minimum size=0.1cm](v4) at (-8.5,1.5) [label=above:$v_{3}$]{};

\draw(v1)--(v2);
\draw(v1)--(v3);
\draw(v3)--(v4);
\draw(v4)--(v2);

\node (G) at (-7.75,-1.5) [label=above:$G$]{};
\node (G1) at (0,-1.5) [label=above:$G'$]{};
    
 \end{tikzpicture}
 \centering
     \caption{Example of the construction in Theorem \ref{twtheorem}.}
     \label{w1hard}
 \end{figure}

 \proof Let $I=(G,k)$ be an instance of  {\sc Exact Defensive Alliance}.
 We construct an instance $I'=(G',k')$ of {\sc Locally Minimal Defensive Alliance} 
 as follows.  See Figure \ref{w1hard}, which provides an illustration of the construction. 
 The construction 
of $G'$ starts with $G':=G$ and then add some new vertices and edges.  Without loss of generality, we can assume that $k < \frac{n}{2} -1$.  First, we introduce a set 
$A=\{a_0, a_{1},\ldots,a_{n-2k}\}$ of $n-2k+1$ new vertices and a new vertex $b_0$ into $G'$.
 For every vertex $a \in A \setminus \{a_0\}$, 
 we introduce a set $V_{a}=\{v_{a}^{1},v_{a}^{2},\ldots,v_{a}^{2n^{3}}\}$ of $2n^3$ vertices into 
 $G'$ and make them adjacent to $a$.  
 Make $a_0$ adjacent to $a_i$ for every $1\leq i\leq n-2k$. Finally, we make $a_{0}$ and $b_{0}$ adjacent to every vertex of $G$. 
 This completes the construction of graph $G'$. 
 We set $k'=(n-k+1)+(n^{3}-1)(n-2k)$. Let  $D$ be a set of vertices such that $V(G)\setminus D$ is a collection of stars. One can easily see that deletion of  $D \cup \{a_{0},b_{0}\}$ from $G'$ results in a collection of stars.

 \par
 To prove the correctness of the reduction, we claim that $G$ has a defensive alliance of size exactly $k$ if and only if $G'$ admits a locally minimal defensive alliance 
 of size at least $k'$. Assume first that $G$ has a defensive alliance of size
 exactly $k$.
 We claim that $S' =  S \cup A \bigcup\limits_{a \in A\setminus \{a_{0}\}} \bigcup\limits_{i=1}^{n^{3}-1} \{v_{a}^{i}\}$ is a locally minimal defensive alliance of size $k'$ in $G'$. Clearly $|S'|\geq k'$. 
 %We claim that $D = S \cup A \cup C$ is a locally minimal defensive alliance in $G'$. 
 Let $x$ be an arbitrary element of $S'$.
 
 \noindent{\it Case 1.} If $x$ is an element of $S$, then we show that it is protected in $S'$. The only new neighbour of $x$ in $G'$ that is part of 
 $S'$ is $a_0$;  the only new neighbour of $x$ in $G'$ that is outside
 $S'$ is $b_0$. Thus the number of defenders and attackers of $x$ in $G'$ increase by 1.
 Therefore, as $x$ was protected in $S$, it remains protected in $S'$.      
  
\noindent{\it Case 2.} If  $x\in \bigcup\limits_{a \in A\setminus \{a_{0}\}} {V_a}$, then the only neighbour of $x$ in $G'$ is in $S'$,
so $x$ can trivially defend itself. 

\noindent{\it Case 3.} If $x=a_0$, the defenders of $x$ consist of all elements of $A\setminus \{a_0\}$ and all elements of $S$. The attackers of 
$x$ consist of all elements of $V(G)\setminus S$. 
Hence $x$ has $n-k$ defenders in  $G'$; and $x$ has $n-k$ attackers in $G'$. 
This shows that $a_0$ is marginally protected. 

\noindent{\it Case 4.} If $x\in A\setminus \{a_0\}$, then 
the defenders of $x$ consist of $n^3-1$ elements of $V_x$ and $a_0$; and the 
attacker of $x$ consist of $n^3+1$ elements of $V_x$. Hence $x$, including itself,
has $n^3+1$ defenders in  $G'$; and $x$ has $n^3+1$ attackers in $G'$. 
This shows that $x$ is marginally protected.
\par As $|S'|>1$ and every vertex of $S'$ has a marginally protected neighbour,  $S'$ is a locally minimal defensive alliance in $G'$. \\

 \par To prove the reverse direction of the equivalence, suppose $G'$ has a locally minimal defensive alliance $S'$ of size at least $k'$. 
 \par First, we claim $A \subseteq S'$. 
 To show this, we first observe that if a degree 1 vertex is in some locally minimal defensive alliance of size at least $2$ then its only neighbour must be marginally protected. Note that $v_a^i$ is a degree 1 vertex.
 Therefore, if  $v_a^i \in V_a$ is present in $S'$ for some $i$ then $a$ has to be in $S'$ and also $a$ has to be marginally protected in $S'$.
 Furthermore $a$ must also have a marginally protected neighbour in $S'$. Since all the vertices in $V_{a} \cap S'$ are strongly protected, the only other 
 neighbour $a_{0}$ of $a$ must be inside $S'$ and $a_0$ must be marginally protected.
 As $a$ is marginally protected, we have $|S' \cap V_{x}| = n^{3}-1$.  
 Otherwise, if $a \not\in S'$ then $V_{a} \cap S' = \emptyset$.
 This would imply that $|S'|<k'$ as for all $x \in A\setminus \{a_{0}\}$, we have  $|S' \cap V_{x}| \leq n^{3}-1$. 
 Therefore we get that $A \subseteq S'$ and all the vertices in $A$ are marginally protected in $S'$. This proves the claim. 
 Note that  $a_{0}$ is marginally protected in $S'$ and $A \subseteq S'$ imply
 that $|S' \cap V(G)|=k$.  
 \par We now claim that $b_0 \not\in S'$. Assume for the sake of contradiction
 that $b_0 \in S'$.  Then  $b_0$, including itself,  has $k+1$ defenders and 
 $n-k$ attackers in $G'$. As $k < \frac{n}{2} -1$, $b_0$ is not protected in 
 $S'$, a contradiction. 
  
  \par Finally we claim that $S=S' \cap V(G)$ is a defensive alliance of size exactly $k$. It is clear that $|S|=k$.  For $S\subseteq V(G)$ and $u\in V(G)$, $d_S(G,u)$ denotes the 
  number of neighbours that $u$ has in $S$ and $d_{S^c}(G,u)$ denotes the 
  number of neighbours that $u$ has outside $S$ of $G$.  Note that for each $u\in S$,
  $d_S(G,u)=d_{S'}(G',u)-1$ and $d_{S^c}(G,u)=d_{{S'}^c}(G',u)-1$  As $S'$ is a defensive alliance in $G'$, we have $d_{S'}(G',u)+1\geq d_{{S'}^c}(G',u)$
 for all $u\in S$. This implies $d_{S}(G,u)+1\geq d_{{S}^c}(G,u)$
 for all $u\in S$.  This shows that $S$ is a defensive alliance.  This completes the proof. \qed\\
%\subsection{W[1]-hardness parameterized by treewidth}

 Clearly stars are trivially acyclic. 
 Moreover, it is easy to verify that stars have 
 pathwidth \cite{Kloks94} and treedepth \cite{Sparsity} at most two, which implies:
 
\begin{theorem}
 {\sc Locally Minimal Defensive Alliance}  
 is W[1]-hard when parameterized by any of the following parameters:
 \begin{itemize}
     \item the feedback vertex set number,
     \item the treewidth and clique width of the input graph,
     \item the pathwidth and treedepth of the input graph.
 \end{itemize}
\end{theorem}

 \section{XP algorithm parameterized by treewidth}\label{treewidthsection}
 This section presents an  XP-algorithm
  for {\sc Locally Minimal Defensive Alliance}  problem parameterized by 
  treewidth.   We prove the following theorem:
  \begin{theorem}\label{treewidth}
  Given an $n$-vertex graph $G$ and its nice tree decomposition $T$ of width at most $k$, the 
  size of a maximum locally minimal defensive alliance of $G$ can be computed in 
  $O({18}^kn^{4k+10})$ time. 
  \end{theorem}  
  
  %In this subsection, we prove the following theorem:
 % \begin{theorem}\label{treewidth}
 % Given an $n$-vertex graph $G$ and its nice tree decomposition $T$ of width at most $k$, the 
  %size of a minimum defensive alliance of $G$ can be computed in $O(2^kn^{2k+4})$ time. 
 % \end{theorem}
  
  Let $(T, \{X_{t}\}_{t \in V(T)})$ be a nice tree decomposition rooted at node $r$ 
  of the input graph $G$.  For a node $t$ of $T$, let $V_t$ be the union of all
bags present in the subtree of $T$ rooted at $t$, including $X_t$. We denote by $G_t$ the subgraph of $
G$ induced by $V_t$.  Here we distinguish not only if a vertex is in the solution or not, but if it is in the solution we also distinguish  if it is marginally protected or not. 
A {\it coloring} of bag $X_t$ is a mapping $f:X_t\rightarrow \{b,w,r\}$ assigning three
different colours to vertices of  the bag.  We give intuition  behind the  three colours.
\begin{itemize}
\item {\bf White}, represented by $w$. The meaning is that all white vertices have to be contained in the partial solution in $G_t$.

    \item {\bf Black}, represented by $b$. The meaning is that all black vertices have to be contained  in 
    the partial solution in $G_t$; additionally, all black vertices must be marginally 
    protected in the final solution. 
    %A vertex is colored $b$ if we want it inside the solution and also marginally protected.
     
    %A vertex is in state $w$ if  we want it inside the solution.
    \item {\bf Red}, represented by $r$. The meaning is that all red vertices are not contained in the partial solution in $G_t$.
    \end{itemize}
For a node $t$, there are $3^{|X_t|}$ colourings $X_t$. Now, for each node $t$ in $T$, we construct a table $dp_t(f, {\bf p},{\bf a}, {\bf v}, \alpha, \pi, \beta, \beta^*, \gamma) \in \{\text{true, false}\}$ where $f$ is a colouring 
of $X_t$, ${\bf p}$ is a vector of length   $n$ such that 
 \[{\bf{p}}(i) = 
            \begin{cases}
                \text{0 or 1}  &\quad\text{if $v_i \in X_t$ and $f(v_i)\in \{b,w\}$}\\
                \text{$\star$} &\quad\text{otherwise;}\\
            \end{cases}   
\] ${\bf a}$ and ${\bf v}$ are vectors of length $n$, and their 
$i$th coordinates are positive only if $v_i$ is in $X_t$ and it is coloured $b$ or $w$;  $\alpha,\pi, \beta, \beta^*$ and $\gamma$ are integers  between $0$ to $n$. 
% For a coloring $f$ of $X_t$ ${\bf p}$ is a vector of length at most  $k$ with entry corresponding to a vertex in $X_{t}$ colored black or white by $f$ is either zero or one,${\bf n}$ and ${\bf q}$ are vectors of length $n$, and its  $i$th coordinate is positive only if $v_i\in A$; $\alpha$ is a integer between $0$ and $n$. The parameters $a,\alpha, \beta, \gamma$ and $\delta$ are integers taking value between $0$ to $n$. 
We set 
$dp_t(f, {\bf p},{\bf a}, {\bf v}, \alpha, \pi, \beta, \beta^*, \gamma)=$ true if and only if there exists  a set $A_{t}\subseteq V_{t}$ such that
%whose vertices are colored either black or white and agrees with coloring $f$ of $X_{t}$. Also if the $i$th entry of vector ${\bf p}$ corresponding to vertex $v_{i}\in A_{t}$ is $1$ then $f^{-1}(v_{i}) \in \{b,w\}$ and $v_{i}$ has a black neighbour in $A_{t}$. Also 
\begin{enumerate}
      \item $\alpha= |A_t|=|\{ v \in V_{t} : f(v)\in \{b,w\} \}|$
      \item $f^{-1}\{b,w\}=A_t\cap X_t=A$, which is the set of vertices of 
      $X_t$ colored black or white. 
      \item  the $i$th coordinate of vector ${\bf p}$ is 
      \[ {\bf p}(i) = 
            \begin{cases}
                \text{1} &\quad\text{if $v_i\in X_t$, $f(v_i)\in \{b,w\}$
                and $v_i$ has a black neighbour in $A_t$}\\
                \text{$0$} &\quad\text{if $v_i\in X_t$, $f(v_i)\in \{b,w\}$
                and $v_i$ has no black neighbours in $A_t$}\\
                \text{$\star$}  &\quad\text{otherwise}\\
                \end{cases}   
\]      
    
      \item  the $i$th coordinate of vector ${\bf a}$ is
      \[ {\bf a}(i) = 
            \begin{cases}
                \text{$d_{A_t}(v_i$)} &\quad\text{if $v_i\in X_t$ and $f(v_i)\in \{b,w\}$}\\
                \text{$0$} &\quad\text{otherwise}\\
            \end{cases}   
\] That is, ${\bf a}(i)$ denotes the number of neighbours of vertex $v_i$ in $A_t$  if $v_i\in X_t$ and $f(v_i)\in \{b,w\}$.
%where $A_{t} = \{v \in V_{t} : f(v)\in \{b,w\} \}$.
\item  the $i$th coordinate of vector ${\bf v}$ is
      \[ {\bf v}(i) = 
            \begin{cases}
                \text{$d_{V_t}(v_i$)} &\quad\text{if $v_i\in X_t$ and $f(v_i)\in \{b,w\}$}\\
                \text{$0$} &\quad\text{otherwise.}\\
            \end{cases}   
\]
That is, ${\bf v}(i)$ denotes the number of neighbours of vertex $v_i$ in $V_t$  if $v_i\in X_t$ and $f(v_i)\in \{b,w\}$.

\item $\pi$ is the number of vertices $v \in A_t$ that are protected, that is, 
     $d_{A_t}(v)\geq \frac{d_G(v)-1}{2}$.
\item  $\beta$ is the number of black vertices in $A_t$.
\item  $\beta^*$ is the number of black vertices $v$ in $A_t$
 such that $N(v)\subseteq V_{t}$  and $d_{A_{t}}(v)= \lceil \frac{d_{G}(v)-1}{2} \rceil$.
 Thus $\beta^*$ is the number of black vertices $v$  in $A_t$ that are marginally 
 protected when all its neighbours are
 introduced in $G_t$.  The intuition here is that we want every back vertex to be 
 marginally protected when all its neighbours are introduced. 
 \item  $\gamma$ is the number of vertices in $A_t$ who has a black neighbour. In other words, $\gamma$ is the number of {\it good} vertices in $A_t$.
\end{enumerate}

%Inthefollowing,we compute all entriesdp(t,n,d,w,s,m)in a bottom-up manner.ThereareO(n·(n+1)c ·2c ·(2n)c · (2n)c ·(n+1)c)=O(n4c+1)possibletuples(t,n,d,w,s,m).Thus,toproveTheorem3.1,itsufficestoshowthateachentry dp(t , n, d, w, s, m) can be computed in time O(n4c ) assuming that the entries for the children of t are already computed.

\noindent We compute all entries $dp_t(f, {\bf p},{\bf a}, {\bf v}, \alpha, \pi, \beta, \beta^*, \gamma)$ in a bottom-up manner.
  Since $tw(T)\leq k$, there are  $ O(3^k \cdot 2^k\cdot n^k \cdot  n^k \cdot (n+1)^5)=O(6^kn^{2k+5})$ possible tuples $(f, {\bf p},{\bf a}, {\bf v}, \alpha, \pi, \beta, \beta^*, \gamma)$.  Thus, to prove Theorem \ref{treewidth}, it suffices to show 
  that each entry $dp_t(f,{\bf p, a, v}, \alpha, \pi, \beta, \beta^*, \gamma)$ can be computed in 
  $O(3^kn^{2k+5})$ time, assuming that the entries for the children of $t$ are already
  computed. 
  
  \begin{lemma}
      For a leaf node $t$, $dp_t(f,{\bf p, a, v}, \alpha,\pi,\beta, \beta^*, \gamma)$ can be computed in O(1) time.  
      \end{lemma}
 \proof  For leaf node $t$ we have that $X_t=\emptyset$. Thus 
$dp_t(f,{\bf p, a, v}, \alpha,\pi,\beta, \beta^*, \gamma)$ is true if and only if $f=\emptyset$, ${\bf p=0}$,
  ${\bf a=0}$, ${\bf v=0}$, $\alpha=0$, $\pi=0$, $\beta=0$, $\beta^*=0$ and $\gamma=0$. These conditions can be checked in 
  $O(1)$ time. \qed

  \begin{lemma}
      For an introduce node $t$, $dp_t(f,{\bf p, a, v}, \alpha,\pi,\beta, \beta^*, \gamma)$ can be computed in O(1) time.  
      \end{lemma} 
      
\proof Suppose $t$ is an introduce node with child $t^{\prime}$ such
 that $X_t=X_{t^{\prime}} \cup \{v_i\}$ for some $v_i\notin X_{t^{\prime}}$. 
 Let $f$ be any coloring of $X_t$. We consider three cases:\\
 
 \noindent{\it Case (i):} Let $f(v_i)=r$. 
 In this case $dp_t(f,{\bf p, a, v}, \alpha,\pi,\beta, \beta^*, \gamma)$ is true if and only if  $dp_{t^{\prime}}(f|_{X_{t'}},{\bf p}, {\bf a}, {\bf v'}, \alpha,\pi,\beta, \beta^*, \gamma)$ is true where
\item 
 \[ {\bf v}(j) = 
            \begin{cases}
                \text{${\bf v}'(j)+1$} & \quad\text{if $j\neq i$, $v_j\in X_t$, $f(v_j)\in \{b,w\}$ and $v_j\in N_{X_t}(v_i)$}\\
                
%                \text{$|N_{X_t}(v_i)|$} &\quad\text{ if $j=i$}\\
                \text{${\bf v}'(j)$} &\quad\text{otherwise}\\
                \end{cases}   
\]

\noindent{\it Case (ii):} Let $f(v_i) =b$. 
Here $dp_t(f,{\bf p},{\bf a}, {\bf v}, \alpha, \pi, \beta, \beta^*, \gamma)$ is true if and only if there exist a tuple $(f',{\bf p'}, {\bf a'}, {\bf v'}, \alpha',\pi',\beta',\beta^{*'}, \gamma')$ such that 
$dp_{t^{\prime}}(f^{\prime}, {\bf p'}, {\bf a^{\prime}}, {\bf v'}, \alpha^{\prime},  \pi^{\prime},\beta^{\prime}, \beta^{*'},\gamma^{\prime})$=true, where 
\begin{enumerate}
\item $f|_{X_{t} \setminus \{ v_{i}\}}=f^{\prime}|_{X_{t'}}$;
\item 
\[{\bf{p}}(j) = 
            \begin{cases}
                \text{1}  &\quad\text{if  $v_j\in X_t$, $f(v_j)\in \{b,w\}$ and 
                $v_j\in N(v_i)$ }\\
                \text{${\bf p'}(j)$} &\quad\text{otherwise}\\
            \end{cases}   
\]

%$p(j)=p'(j)$ if $f(v_{j})=\{b,w\}$ and $v_{j}\not\in N(v_{i})$;  $p(j)=1$ if $f(v_{j})=\{b,w\}$ and $v_{j}\in N(v_{i})$ ;  
\item  \[ {\bf a}(j) = 
            \begin{cases}
                \text{${\bf a}'(j)+1$} & \quad\text{if $j\neq i$, $v_j\in X_t$, $f(v_j)\in \{b,w\}$ and $v_j\in N_{X_t}(v_i)$}\\
                
                \text{$|N_A(v_i)|$} &\quad\text{if $j=i$}\\
                \text{${\bf a}'(j)$} &\quad\text{otherwise}\\
                \end{cases}   
\]
where $A=A_t\cap X_t$.

%$n(j)=n^{\prime}(j)+1$, if $v_j\in N(v_i)$ and $f(v_{j})\in \{b,w\}$, otherwise $n(j)=n^{\prime}(j)$; $n(i)=d_{A}(v_{i})$ where $A = \{v \in X_{t} ~|~ f(v)\in \{b,w\}\}$;
\item 
 \[ {\bf v}(j) = 
            \begin{cases}
                \text{${\bf v}'(j)+1$} & \quad\text{if $j\neq i$, $v_j\in X_t$, $f(v_j)\in \{b,w\}$ and $v_j\in N_{X_t}(v_i)$}\\
                
                \text{$|N_{X_t}(v_i)|$} &\quad\text{ if $j=i$}\\
                \text{${\bf v}'(j)$} &\quad\text{otherwise}\\
                \end{cases}   
\]

%$q(j)=q^{\prime}(j)+1$, if $v_j\in N(v_i)$ and $f(v_{j})\in \{b,w\}$, otherwise $q(j)=q^{\prime}(j)$; $q(i)=d_{A}(v_{i})$ where $A = \{v \in X_{t} ~|~ f(v)\in \{b,w\}\}$;
\item $\alpha=\alpha^{\prime}+1$;
\item $\pi=\pi^{\prime} +l$; here $l$ is the cardinality of the set $$\Big\{ v_j\in X_t~|~f(v_j)\in \{b,w\},  {\bf a}^{\prime}(j) < \frac{d_G(v_j)-1}{2} ; {\bf a}(j) \geq  \frac{d_G(v_j)-1}{2}  \Big\}.$$
That is, to compute $\pi$ from $\pi^{\prime}$ we need to add the number $l$ of vertices 
$v_j \in X_t$
which are not protected in  $X_{t'}$
but protected in $X_t$.
\item $\beta = \beta'+1$;
\item 
$\beta^* = \beta^{*'} +  \delta $ \\
where  $\delta$ is the number of black vertices $v_j \in X_t$ such that
$ {\bf a'}(j) \neq  \lceil \frac{d_G(v_j)-1}{2} \rceil \ \text{or} \ {\bf v'}(j) \neq d(v_{j})$ but it satisfies the conditions ${\bf a}(j) =  \lceil \frac{d_G(v_j)-1}{2} \rceil$ and  $ {\bf v}(j)= d(v_j)$.

\item $\gamma = \gamma'+|\{ v_{j}\in A ~|~ {\bf p}(j)=1$ but ${\bf p}'(j)=0\}|$.
\end{enumerate}

\noindent{\it Case (iii):} Let $f(v_i) =w$. 
Here $dp_t(f,{\bf p},{\bf a}, {\bf v}, \alpha, \pi, \beta, \beta^*, \gamma)$ is true if and only if there exist a tuple $(f',{\bf p'}, {\bf a'}, {\bf v'}, \alpha',\pi',\beta',\beta^{*'}, \gamma')$ such that 
$dp_{t^{\prime}}(f^{\prime}, {\bf p'}, {\bf a^{\prime}}, {\bf v'}, \alpha^{\prime},  \pi^{\prime},\beta^{\prime}, \beta^{*'},\gamma^{\prime})$=true, where

\begin{enumerate}
\item $f|_{X_{t} \setminus \{ v_{i}\}}=f'$;

\item 
\[{\bf{p}}(j) = 
            \begin{cases}
               % \text{1}  &\quad\text{if  $v_j\in X_t$, $f(v_j)\in \{b,w\}$ and 
               % $v_j\in N(v_i)$ }\\
                \text{${\bf p'}(j)$} &\quad\text{if  $j\neq i$}\\
                \text {1} &\quad\text{if $j=i$ and $v_i$ has a black neighbour in $X_t$}\\
                $0$ &\quad\text{if $j=i$ and $v_i$ has no black neighbours in $X_t$}
            \end{cases}   
\]

%$p(j)=p'(j)$ for all $j\neq i$ such that  $f(v_{j})=\{b,w\}$; $p(i)=1$ if and only if $(v_{i},v_{j})\in E(G)$ where $f(v_{j})=b$ for some $j$;\item  \[ {\bf n}(j) = 

\item  \[ {\bf a}(j) = 
            \begin{cases}
                \text{${\bf a}'(j)+1$} & \quad\text{if $j\neq i$, $v_j\in X_t$, $f(v_j)\in \{b,w\}$ and $v_j\in N_{X_t}(v_i)$}\\
                
                \text{$|N_A(v_i)|$} &\quad\text{ if $j=i$}\\
                \text{${\bf a}'(j)$} &\quad\text{otherwise}\\
                \end{cases}   
\]
where $A=A_t\cap X_t$.

\item 
 \[ {\bf v}(j) = 
            \begin{cases}
                \text{${\bf v}'(j)+1$} & \quad\text{if $j\neq i$, $v_j\in X_t$, $f(v_j)\in \{b,w\}$ and $v_j\in N_{X_t}(v_i)$}\\
                
                \text{$|N_{X_t}(v_i)|$} &\quad\text{ if $j=i$}\\
                \text{${\bf v}'(j)$} &\quad\text{otherwise}\\
                \end{cases}   
\]

%\item $n(j)=n^{\prime}(j)+1$, if $v_j\in N(v_i)$ and $f(v_{j})\in \{b,w\}$, otherwise $n(j)=n^{\prime}(j)$; $n(i)=d_{A}(v_{i})$ where $A = \{v \in X_{t} ~|~ f(v)\in \{b,w\}\}$;
%\item $q(j)=q^{\prime}(j)+1$, if $v_j\in N(v_i)$ and $f(v_{j})\in \{b,w\}$, otherwise $q(j)=q^{\prime}(j)$; $q(i)=d_{A}(v_{i})$ where $A = \{v \in X_{t} ~|~ f(v)\in \{b,w\}\}$;
\item $\alpha=\alpha^{\prime}+1$;
\item $\pi=\pi^{\prime} +l$; here $l$ is the cardinality of the set $$\Big\{ v_j\in X_t~|~f(v_j)\in \{b,w\},  {\bf a}'(j) < \frac{d_G(v_j)-1}{2} ; {\bf a}(j) \geq  \frac{d_G(v_j)-1}{2}  \Big\}.$$
%$\alpha=\alpha^{\prime} +l$; here $l$ is the cardinality of the set $$\Big\{ f(v_j)\in \{b,w\}~|~ n^{\prime}(j) < \frac{d_G(v_j)-1}{2} ; n(j) \geq  \frac{d_G(v_j)-1}{2}  \Big\}.$$
That is, to compute $\pi$ from $\pi^{\prime}$ we need to add the number $l$ of vertices 
$v_j \in X_t$
which are not protected in  $X_{t'}$
but protected in $X_t$.

\item $\beta = \beta'$;
\item 
$\beta^* = \beta^{*'} +  \delta $ \\
where  $\delta$ is the number of black vertices $v_j \in X_t$ such that
$ {\bf a'}(j) \neq  \lceil \frac{d_G(v_j)-1}{2} \rceil \ \text{or} \ {\bf v'}(j) \neq d(v_{j})$ but it satisfies the conditions ${\bf a}(j) =  \lceil \frac{d_G(v_j)-1}{2} \rceil$ and  $ {\bf v}(j)= d(v_j)$.

\item $\gamma = \gamma'+1$ if $v_i$ is adjacent to a vertex in $X_t$ which is coloured black;  otherwise $\gamma=\gamma^{\prime}$.
\end{enumerate}

\noindent For introduce node $t$, $dp_t(f,{\bf p},{\bf a}, {\bf v}, \alpha, \pi, \beta, \beta^*, \gamma)$ can be computed in $O(1)$ time as  there is only one candidate of such tuple $(f',{\bf p'}, {\bf a'}, {\bf v'}, \alpha',\pi',\beta',\beta^{*'}, \gamma')$.  \qed\\

 \begin{lemma}
      For a forget node $t$, $dp_t(f,{\bf p, a, v}, \alpha,\pi,\beta, \beta^*, \gamma)$ can be computed in $O(n)$ time.  
      \end{lemma} 
\proof Suppose $t$ is a forget node with child $t^{\prime}$ such that $X_t=X_{t^{\prime}}\setminus \{v_i\}$ for some 
$v_i\in X_{t^{\prime}}$. Here $dp_t(f,{\bf p},{\bf a}, {\bf v}, \alpha,\pi,\beta, \beta^*, \gamma)$ is true if and only if
$dp_{t'}(f',{\bf p'},{\bf a'}, {\bf v'}, \alpha',\pi',\beta', \beta^{*'},\gamma')$ is true, where\\

\begin{enumerate}
\item $f'=f_{v_i\rightarrow b}, f_{v_i\rightarrow w}$ or $f_{v_i\rightarrow r}$.
\item \[{\bf{p}}(j) = 
            \begin{cases}
               % \text{1}  &\quad\text{if  $v_j\in X_t$, $f(v_j)\in \{b,w\}$ and 
               % $v_j\in N(v_i)$ }\\
                \text{${\bf p'}(j)$} &\quad\text{if  $j\neq i$}\\
               % \text {1 or 0} &\quad\text{if $j=i$ and $f'(v_i)=\{b,w\}$ }\\
                \text{$\star$} &\quad\text{if $j=i$}
            \end{cases}   
\]

%\item $p(j)=p^{\prime}(j) $ for all $j \neq i$;
\item ${\bf a}(j)={\bf a'}(j) $ for all $j \neq i$ and ${\bf a}(i)=0$;
\item ${\bf v}(j)={\bf v'}(j) $ for all $j \neq i$ and ${\bf v}(i)=0$;
\item $\alpha=\alpha'$;
\item $\pi=\pi^{\prime}$;
\item $\beta=\beta^{\prime}$;
\item $\beta^*=\beta^{*'}$;
\item $\gamma=\gamma^{\prime}$.
\end{enumerate}
There are $n+1$ choices for ${\bf a'}(i)$ and ${\bf v'}(i)$ each.  Thus the lemma follows as there are $O(n)$ candidates of such tuples 
$(f',{\bf p'},{\bf a'}, {\bf v'}, \alpha',\pi',\beta', \beta^{*'},\gamma')$. 
This completes the proof of the lemma. \qed\\

\begin{lemma}
      For a join node $t$, $dp_t(f,{\bf p, a, v}, \alpha,\pi,\beta, \beta^*, \gamma)$ can be computed in $O(3^kn^{3k+5})$ time.  
      \end{lemma} 
\proof  Suppose $t$ is a join node with children $t_1$ and $t_2$ such that $X_t=X_{t_1}=X_{t_2}$. 
%Let $A$ be any subset of $X_t$. 
Then $dp_t(f,{\bf p}, {\bf a}, {\bf v}, \alpha,\pi,\beta, \beta^*, \gamma)$ is true 
if and only if there exist $(f_{1},{\bf p}_{1}, {\bf a}_{1}, {\bf v}_{1},\alpha_{1},\pi_{1},\beta_{1}, \beta^*_1, \gamma_{1})$ and 
$(f_{2},{\bf p}_{2}, {\bf a}_{2}, {\bf v}_{2}, \alpha_{2},\pi_{2},\beta_{2}, \beta_2^*, \gamma_{2})$ such that
$dp_{t_1}(f_{1},{\bf p}_{1}, {\bf a}_{1}, {\bf v}_{1},\alpha_{1},\pi_{1},\beta_{1}, \beta_1^*, \gamma_{1})$  and $dp_{t_2}(f_{2},{\bf p}_{2}, {\bf a}_{2}, {\bf v}_{2}, \alpha_{2},\pi_{2},\beta_{2}, \beta_2^*, \gamma_{2})$ 
are true, where

\begin{enumerate}
\item $f=f_1=f_2$;
\item ${\bf p}(i)=1$ if ${\bf p}_{1}(i)=1$ or ${\bf p}_{2}(i)=1$;
\item ${\bf a}(i)={\bf a}_1(i)+{\bf a}_2(i)-d_A(v_i)$ for all $v_{i}\in A$, and ${\bf a}(i)=0$ if $v_{i} \notin A$ where $A = \{v \in X_{t} ~|~ f(v)\in \{b,w\}\}$;
\item ${\bf v}(i)={\bf v}_1(i)+{\bf v}_2(i)-d_{X_t}(v_i)$ for all $v_{i}\in A$, and ${\bf v}(i)=0$ if $v_{i} \notin A$;
\item $\alpha=\alpha_1+\alpha_2-|A|$;
\item  $\pi=\pi_1+\pi_2-l_{1}+l_{2} $;\\ where $l_{1}$ is the cardinality of the set 
$$ \Big\{ v_j\in A~|~ {\bf a}_1(j) \geq \frac{d_G(v_i)-1}{2}; ~{\bf a}_2(j) \geq \frac{d_G(v_i)-1}{2} \Big\} $$  and $l_{2}$ is 
the cardinality of the set $$ \Big\{ v_j\in A~|~ {\bf a}_1(j) < \frac{d_G(v_i)-1}{2}; ~{\bf a}_2(j) < \frac{d_G(v_i)-1}{2}; ~{\bf a}(j) \geq \frac{d_G(v_i)-1}{2}  \Big\}. $$
To compute $\pi$ from $\pi_1+\pi_2$, we need to subtract the number of those $v_j$ which 
are protected in both the 
branches and add the number of vertices $v_j$ which are not protected in either of the branches $t_1$ and $t_2$
but protected in $t$. 

\item $\beta= \beta_{1}+\beta_{2}- |\{v\in A ~|~ f(v)=b\}|; $

\item $\beta^* = \beta_{1}^*+ \beta_{2}^* +\delta_1+\delta_2-\delta_{12}$.\\
Here $\delta_1$ is the number of black vertices $v_j$ in $X_t$ such that
$ {\bf a}_{1}(j) \neq  \lceil \frac{d_G(v_j)-1}{2} \rceil \ \text{or} \ {\bf v}_{1}(j) \neq d(v_{j})$ but it satisfies the conditions ${\bf a}(j) =  \lceil \frac{d_G(v_j)-1}{2} \rceil$ and  $ {\bf v}(j)= d(v_j)$. Similarly, $\delta_2$ is the number of black vertices $v_j$ in $X_t$ such that either 
$ {\bf a}_{2}(j) \neq  \lceil \frac{d_G(v_j)-1}{2} \rceil \ \text{or} \ {\bf v}_{2}(j) \neq d(v_{j})$ but it satisfies the conditions ${\bf a}(j) =  \lceil \frac{d_G(v_j)-1}{2} \rceil$ and  $ {\bf v}(j)= d(v_j)$. Finally $\delta_{12}$ is the number of black vertices 
$v_j$ in $X_t$ such that   
$ {\bf a}_{1}(j) \neq  \lceil \frac{d_G(v_j)-1}{2} \rceil\ \text{or} \ {\bf v}_{1}(j) \neq d(v_{j})$, and $ {\bf a}_{2}(j) \neq  \lceil \frac{d_G(v_j)-1}{2} \rceil \ \text{or} \ {\bf v}_{2}(j) \neq d(v_{j})$ but it satisfies the conditions ${\bf a}(j) =  \lceil \frac{d_G(v_j)-1}{2} \rceil$ and  $ {\bf v}(j)= d(v_j)$.

\item $\gamma = \gamma_{1} + \gamma_{2} - |\{v\in A ~|~ {\bf p}_{1}(v)={\bf p}_{2}(v)=1\}|.$
\end{enumerate}
For join node $t$, there are at most $3^k$ possible pairs for 
$({\bf p_1, p_2 })$ as 
$({\bf p_1}(i), {\bf p_2 }(i)) \in \{(1,0), (0,1), (1,1)\}$ when ${\bf p}(i)=1$ and
$({\bf p_1}(i), {\bf p_2 }(i)) = (0,0)$ when ${\bf p}(i)=0$; there are  $n^k$ possible pairs for $(\bf {a_1,a_2})$ as ${\bf a_2}$ is uniquely determined by ${\bf a_1}$; there are  $n^k$ possible pairs for $(\bf {v_1,v_2})$ as ${\bf v_2}$ is uniquely determined by ${\bf v_1}$; $n+1$ possible pairs for $(\alpha_1,\alpha_2)$; $n+1$ possible pairs for $(\pi_1,\pi_2)$; $n+1$ possible pairs for $(\beta_1,\beta_2)$; $n+1$ possible pairs for $(\beta^*_1,\beta^*_2)$; and 
$n+1$ possible pairs for $(\gamma_1,\gamma_2)$. In total, there are
$O(3^k n^{2k+5})$ candidates, and each of them can be checked in $O(1)$ time. Thus, for join node $t$,  
$dp_t(f, {\bf p}, {\bf a}, {\bf v}, \alpha,\pi,\beta, \beta^*, \gamma)$ can be computed in $O(3^kn^{2k+5})$ time.\qed\\
 
 \noindent At the root node $r$, we look at all records such that $dp_r(\emptyset, \emptyset,\emptyset, \emptyset,  \alpha,\pi,\beta, \beta^*, \gamma)$= true,
  $\beta = \beta^{*}$ (that is, all black vertices in the solution are marginally protected) and  $\alpha=\pi=\gamma$ (that is, every vertex in the solution is protected and has a black or marginally protected neighbour). The size of a maximum locally minimal defensive alliance is the maximum $\alpha$ satisfying 
 $dp_r(\emptyset, \emptyset,\emptyset, \emptyset,  \alpha,\pi,\beta, \beta^*, \gamma)$= true, $\alpha=\pi=\gamma$ and $\beta = \beta^{*}$. \\\\
 {\bf Remark.} The above algorithm implies that {\sc  Locally Minimal Defensive Alliance} can be solved in polynomial time on trees. It is not difficult to modify the above algorithm  to find a connected locally minimal defensive alliance of maximum size.  This means that one can also get an XP algorithm  for {\sc Connected Locally Minimal Defensive Alliance} parameterized by treewidth.

\section{Conclusion} The main contributions in this paper are 
that {\sc Locally Minimal Defensive Alliance}  
 is W[1]-hard when parameterized by any of the following parameters: feedback vertex set number, treewidth, clique width, pathwidth and treedepth of the input graph, and 
 the problem is XP in treewidth. We also proved that 
{\sc Locally Minimal Defensive Alliance}  is NP-complete on planar graphs and FPT 
when parameterized by neighbourhood diversity.   We gave a 
randomized FPT algorithm for {\sc Exact Connected Locally Minimal Defensive Alliance}.
 We list some nice problems emerge from the results here. 
 The question whether {\sc Locally Minimal Defensive Alliance} is FPT when parameterized by solution size has still remained open.   Noting that the result for neighbourhood diversity implies that the problem is FPT in vertex cover, it would be interesting to consider the parameterized complexity with respect to twin cover. The modular width parameter also appears to be a natural parameter to consider here, and since there are graphs with bounded modular-width and unbounded neighbourhood diversity; we believe this is also an 
interesting open problem.  The parameterized complexity of {\sc Locally Minimal Defensive Alliance } 
remains unsettled  when parameterized by other important 
structural graph parameters like vertex integrity and cluster vertex deletion.

\section*{Acknowledgement} We thank the editor and anonymous reviewers for their
constructive comments and suggestions, which helped us to improve the manuscript. 
%We are grateful to David Manlove for  introducing us to the {\sc $\mathcal{F}$-Free Edge Deletion} problem and  for helpful advice and fruitful discussions.
The first author gratefully acknowledges support from the Ministry of Education, 
 Government of India, under Prime Minister's Research Fellowship Scheme (No. MRF-192002-211). 

\bibliographystyle{abbrv}
\bibliography{bibliography}
\newpage
%\appendix
%\input{appendix}
\end{document}